%% file: paper.tex
\documentclass{article}

\usepackage[utf8]{inputenc}  
\usepackage[T2A]{fontenc}    
\usepackage[russian,english]{babel}  

\usepackage{url}
\usepackage{epsfig}
\usepackage{wrapfig}
\usepackage{listings}
\usepackage{multicol}
\usepackage{color}
\usepackage{amsmath}
\usepackage{amsfonts}
\usepackage{stmaryrd}
\usepackage{listings}
\usepackage{boxedminipage}

\lstdefinelanguage{pp}{%
  numbers=none,
  literate={SPACE}{{\ }}1 {EPSILON}{{$\varepsilon$}}1 
  {*}{{$^\star$}}1 {+}{{$^+$}}1 {?}{{$?$}}1 {??}{{?}}1 {<}{{$\langle$}}1 {<<}{{<}}1 {>}{{$\rangle$}}1 {>>}{{>}}1 {FAKE-INLINE}{{inline}}1,
  keywordstyle=\normalfont\bfseries,
 morekeywords={unfold,fold,inline,extract,abridge,detour,unchain,chain,
 massage,distribute,factor,deyaccify,yaccify,eliminate,introduce,import,vertical,horizontal,equate,rename,
 renameL,renameN,renameS,renameT,rassoc,lassoc,
 add,addV,addH,appear,widen,upgrade,unite,
 remove,removeV,removeH,disappear,narrow,downgrade,
 abstractize,concretize,permute,
 define,undefine,redefine,inject,project,replace,
 designate,unlabel,deanonymize,anonymize,dump,reroot,in},
  columns=fullflexible,
  basicstyle=\tt,
}

\newcommand{\doi}[1]{\url{http://dx.doi.org/#1}}

\newcommand{\backuseq}{$:\equiv$\ }
\newcommand{\backusor}{$\overline{\texttt{or}}$\ }
\newcommand{\crleft}{\ensuremath{\hookleftarrow}}

\newcommand{\linktoxbgf}[1]{\href{http://slps.svn.sourceforge.net/viewvc/slps/topics/grammars/wiki/mediawiki-bnf/#1.xbgf?view=markup}{\texttt{#1.xbgf}}}
\newcommand{\linktoedd}[1]{\href{http://slps.svn.sourceforge.net/viewvc/slps/topics/grammars/wiki/mediawiki-bnf/#1.edd?view=markup}{\texttt{#1.edd}}}
\newcommand{\linktofile}[1]{\href{http://slps.svn.sourceforge.net/viewvc/slps/topics/grammars/wiki/mediawiki-bnf/#1?view=markup}{\texttt{#1}}}

\usepackage[unicode,bookmarks=false,pdfstartview={FitH},%
            colorlinks,linkcolor=blue,urlcolor=blue,citecolor=blue,%
            pdfauthor={Dr. Vadim Zaytsev},
            pdftitle={Foundations for a Grammar Laboratory: Requirements Specification}]{hyperref}
\begin{document}
	\English

\title{MediaWiki Grammar Recovery}

\author{Vadim Zaytsev, \url{vadim@grammarware.net}\\
\href{http://www.cwi.nl/research-groups/Software-Analysis-and-Transformation}{SWAT},
\href{http://www.cwi.nl/}{CWI}, NL}

\date{\today}

\maketitle\sloppy

\section{Introduction}

Wiki is the simplest online database that could possibly work~\cite{leuf_cunningham2001}. It usually takes a form of a website or a webpage
where the presentation is predefined to some extent, but the content can be edited by a subset of users. The editing ideally does not require
any additional software nor extra knowledge, takes place in a browser and utilises a simple notation for markup. Currently there are more than
a hundred of such notations, varying slightly in concrete syntax but mostly providing the same set of features for emphasizing fragments of
text, making tables, inserting images, etc~\cite{WikiEngines}. The most popular notation of all is the one of MediaWiki engine, it is used on
Wikipedia, Wikia and numerous Wikimedia Foundation projects.

In order to facilitate development of new wikiware and to simplify maintenance of existing wikiware, one can rely on methods and tools from
software language engineering. It is a field that emerged in recent years, generalising theoretical and practical aspects of programming
languages, markup languages, modelling languages, data definition languages, transformation languages, query languages, application programming
interfaces, software libraries, etc~\cite{MegaModel,SLE-DSL,Towards,Bridging} and believed to be the successor for the object-oriented
paradigm~\cite{SLE-OOP}. The main instrument of software language engineering is on disciplined creation of new domain specific languages with
emphasis on extensive automation. Practice shows that automated software maintenance, analysis, migration and renovation deliver considerable
benefits in terms of costs and human effort compared to alternatives (manual changes, legacy rebuild, etc), especially on large
scale~\cite{Cordy:2003:CRP:851042.857051,vandenBrand:2000:GCS,NielsPhD}. However, automated methods do require special foundation for their
successful usage.

Wikiware (wiki engines, parsers, bots, etc) is a specific case of grammarware (parsers, compilers, browsers, pretty-printers, analysis and
manipulation tools, etc)~\cite{Towards,Zaytsev-Thesis2010}. The most straightforward definition of grammarware can be of software which input
and/or output must belong to a certain language (i.e., can be specified implicitly or explicitly by a formal grammar). An operational grammar
is needed to parse the code, to get it from a textual form that the programmers created into a specialised generational and transformational
infrastructure that usually utilises a tree-like internal format. In spite of the fact that the formal grammar theory is quite an established
area since 1956~\cite{Chomsky}, the grammars of mainstream programming languages are rarely freely obtainable, they are complex artefacts that
are seen as valuable IT assets, require considerable effort and expertise to compose and therefore are not always readily disclosed to public
by those who develop, maintain and reverse engineer them. A syntactic grammar is basically a mere formal description of what can and what
cannot be considered valid in a language. The most obvious sources for this kind of information are: language documentation, grammarware source
code, international standards, protocol definitions, etc.

However, documentation and specifications are neither ever complete nor error-free~\cite{LDF2011}.
To obtain correct grammars and ensure their quality level, special techniques are needed:
	grammar adaptation~\cite{Adaptation},
	grammar recovery~\cite{Recovery},
	grammar engineering~\cite{Towards},
	grammar derivation~\cite{Tolerant},
	grammar reverse engineering,
	grammar re-engineering,
	grammar archaeology~\cite{GRK},
	grammar extraction~\cite[\S5.4]{Zaytsev-Thesis2010},
	grammar convergence~\cite{Convergence2009},
	grammar relationship recovery~\cite{JLS-journal},
	grammar testing~\cite{Laemmel01:FASE},
	grammar inference~\cite{Inference},
	grammar correction~\cite[\S5.7]{Zaytsev-Thesis2010},
	programmable grammar transformation~\cite{XBGF-manual},
and so on.
The current document is mainly a demonstration of application of such techniques to the MediaWiki BNF grammar that was published as
\cite{MW-BNF-Article-title,MW-BNF-Article,MW-BNF-Noparse-block,MW-BNF-Links,MW-BNF-Magic-links,MW-BNF-Special-block,MW-BNF-Inline-text,MW-BNF-Fundamentals}.

\subsection{Objectives}

The project reported in this document aims at extraction and initial recovery of the MediaWiki grammar.
However, the extracted grammar is not the final goal, but rather a stepping stone to enable the following activities:

\begin{description}
	\item[Parse wiki pages.]
		The current state of Wikipedia is based on a PHP rewriting system that transforms wiki layout directly into HTML~\cite{MW-Parser}.
		However, it can not always be utilised in other external wikiware: for example, future plans of Wikimedia Foundation
		include having an in-browser editor with a WYSIWYG front-end in JavaScript~\cite{MW-Parser-plan}.
		Having an operational grammar means anyone can parse wiki pages more freely with their own technology of choice,
		either directly or by deriving tolerant grammars from the baseline grammar~\cite{Tolerant}.
	\item[Aid wiki migration.]
		The ability to easily parse and transform wiki pages can deliver considerable benefits when migrating
		wiki content from one platform to another~\cite{WikiMigration2011}.
	\item[Validate existing wiki pages.]
		The current state of MediaWiki parser~\cite{MW-Parser} allows users to submit wiki pages that are essentially incorrect:
		they may combine wiki notation with bare HTML, contain unbalanced markup, refer to nonexistent templates.
		This positively affects the user-friendliness of the wiki, but makes some wiki pages possibly problematic.
		Such pages can be identified and repaired with static code analysis techniques~\cite{SCA07}.
	\item[Test existing wiki parsers.]
		There is considerable prior research in the field of grammar-based testing, both
		stochastic~\cite{Maurer90,SirerB1999} and
		combinatorial~\cite{HennessyPower08,Laemmel01:FASE,MalloyPower01,Purdom72,Zaytsev-Thesis2004},
		with important recent advances in formulating coverage criteria and achieving automation~\cite{TestMatch2011,LaemmelS06}.
		These results can be easily reproduced to provide an extensive test data suite containing
		different wiki text fragments to explore every detail specified by the grammar in a fully automated fashion.
		Such test data suites can be used to determine existing parsers' conformance, can help in developing new
		parsers, find problematic combinations that are treated differently by different parsers, etc.
	\item[Improve grammar readability.]
		It is known that the grammar is meant to both define the language for the computer to parse, and describe it
		for the language engineers to understand. However, these two goals are usually conflicting, and more often than not,
		one opts for an executable grammar that is harder to read, than for a perfectly readable one that cannot be used
		in constructing grammarware. Unfortunately, the effort and expertise needed to fully achieve either of them,
		and most language documents contain non-operational grammars~\cite{NeedsGrammarware2005,Zaytsev-Thesis2004,LDF2011}.
		The practice of using two grammars: the ``more readable'' one and the ``more implementable'', adopted in the
		Java specification~\cite{JLS3}, has also proven to be very ineffective and error-prone~\cite{JLS-SCAM2009,JLS-journal}.
	\item[Perform automated adaptation.]
		Grammars commonly need to be adapted in order to be useful and efficient in wide range of circumstances~\cite{Adaptation}.
		Grammar transformation frameworks such as GDK~\cite{GDK}, GRK~\cite{GRK} or XBGF~\cite{XBGF-manual} can be used to
		apply adapting transformations in a safe disciplined way with validation of applicability preconditions and full control
		over the language delta. In fact, some of the transformations can even be generated automatically and applied afterwards.
	\item[Establish inter-grammar relationships.]
		As of today, several MediaWiki notation grammars exist and are available in one form or another:
			in EBNF~\cite{MW-EBNF},
			in ANTLR~\cite{MW-ANTLR}, etc (none of them are fully operational).
		Furthermore, there exist various other wiki notations:
			Creole~\cite{Creole2007},
			Wikidot~\cite{Wikidot}, etc.
		Relationships among all these notations are unknown: they are implicit even when formal grammars actually exist,
		and are totally obscured when the notation is only documented in a manual. A special technique called language
		convergence can help to reengineer such relationships in order to make stronger claims about compatibility and
		expressivity~\cite{Convergence2009,Zaytsev-Thesis2010,LCI2011}.
\end{description}

\subsection{Related work: grammar recovery initiatives}

Most of operational grammars for mainstream software languages are handcrafted, many are not publicly disclosed, few are documented. The first
case reported in detail in 1998 was PLEX (Programming Language for EXchanges), a proprietary DSL for real time embedded software systems by
Ericsson~\cite{PLEX}, a successful application of the same technology to COBOL followed~\cite{COBOL-recovery}. Grammar recovery technique is
not only needed for legacy languages, examples of more modern and presumably more accurately engineered grammars being nontrivially extracted
include C\# in \cite{TooSharp2005} and \cite[\S3]{Zaytsev-Thesis2010} and Java in \cite{JLS-SCAM2009} and \cite{JLS-journal}. The whole
process of MediaWiki grammar extraction is documented by this report, all corrections and refactorings are available online, as is the end
result (under CC-BY-SA license).

\subsection{Related work: Wiki Creole}

Wiki Creole 1.0\footnote{\url{http://wikicreole.org}} is an attempt for engineering an ideal
wiki syntax and a formal grammar for it. While the goal of specifying the wiki syntax with a grammar
is not foreign to us, but the benefits listed in \cite[p.3]{Creole2007} are highly questionable:

\begin{description}
	\item[1. Trivial parser construction.]
		In the paper cited above it is claimed that applying a parser generator is trivial.
		However, the main prerequisite for it is successful grammar adaptation for the
		particular parsing technology~\cite{Adaptation}. A Wiki Creole grammar was specifically geared toward ANTLR,
		and it is a highly sophisticated task to migrate it anywhere if at some point ANTLR use
		is deemed to be undesirable.
		Hence, the result is not reproducible without considerable effort and expertise.
	\item[2. Foundation for subsequent semantics specification.]
		The grammar can certainly serve as a basis for specifying semantics.
		However, the choice of a suitable calculus for such semantics specification is of even more importance.
		Furthermore, syntax definition does not guarantee the absence of ambiguities in semantics,
		or even changes of semantics as a part of language evolution (cf., evolutionary changes of HTML elements).
	\item[3. Improved communication between wikiware developers.]
		The paper claimed that if wiki syntax is specified with a grammar, there can be no different interpretations of it.
		However, it is quite common to have different interpretations (dialects) of even mainstream programming languages,
		plus wiki technology in its current state heavily relies on fault tolerance (somewhat less so in the future when no bare text
		editing should be taking place).
	\item[4. Same rendering behaviour that users rely on.]
		Depending on the browser or the particular gadget that the end user deploys to access the wiki, rendering behaviour
		can be vastly different, and this has nothing to do with the syntax specification.
	\item[5. Simplified syntax extension.]
		It is a very known fact in formal grammar theory~\cite{DragonBook} that grammar classes are not compositional: that is, the result
		of combining two LL(*) grammars (which ANTLR uses) does not necessarily belong to the LL(*) class; we can only
		prove that it will still be context-free~\cite{Chomsky}. In other words, it is indeed easy to specify a syntax extension, but such
		the extended grammar sometimes will not be operational.
		Modular grammars can be deployed in frameworks which use different parsing technologies, such as in Meta-Environment~\cite{ASFSDF-Klint}
		or in Rascal~\cite{Rascal} or in MPS~\cite{DBLP:conf/models/Voelter10}, but not in ANTLR.
	\item[6. Performance predictions.]
		The paper claims that it is easier to predict performance of a parser made with ``well-understood language theory'' than with
		a parser based on regular expressions. However, there are implementation algorithms of regular expressions that demonstrate 
		quadratic behaviour~\cite{CoxRE07}, and ANTLR uses the same technology for matching lookahead anyway, which immediately means
		that their performance is the same.
	\item[7. Discovering ambiguities.]
		It is true that ambiguity analysis is easier on a formal grammar than on the prose, but it is not achieved by ``more rigorous specification
		mechanism'' and even the most advanced techniques of today do not always succeed~\cite{Basten2010}.
	\item[8. Well-defined interchange format.]
		A well designed interchange format between different types of wikiware is a separate effort that should be based on appropriate
		generalisations of many previously existing wiki notations, not on one artificially created one, even if that one is better designed.
\end{description}

In general, Wiki Creole initiative is relevant for us because it can serve as a common grammar denominator later to converge several wiki
grammars~\cite{Convergence2009,LCI2011}, but is neither contributing nor conflicting directly with our grammar recovery project.


\section{Grammar notation}
\label{ceci-n'est-pas-une-bnf}

One of the first steps in grammar extraction is understanding the grammar definition formalism (i.e., the notation) used in the original
artefact to describe the language. In the case of MediaWiki, Backus-Naur form is claimed to be used~\cite{MW-BNF}. Manual cursory examination of
the grammar text
\cite{MW-BNF-Article-title,MW-BNF-Article,MW-BNF-Noparse-block,MW-BNF-Links,MW-BNF-Magic-links,MW-BNF-Special-block,MW-BNF-Inline-text,MW-BNF-Fundamentals}
allows us to identify the following metasymbols in the spirit of \cite{ISO-EBNF} and \cite{Zaytsev-Thesis2010}:

\begin{center}\begin{tabular}{|l|c|}\hline
	Name	& Value\\\hline
	Start grammar symbol			&	\texttt{<source lang=bnf>}\\
	End grammar symbol				&	\texttt{</source>}\\
	Start comment symbol			&	\texttt{/*}\\
	End comment symbol          	&	\texttt{*/}\\
	Defining symbol					&	\texttt{::=}\\
	Definition separator symbol 	&	\texttt{|}\\
	Start nonterminal symbol		&	\texttt{<}\\
	End nonterminal symbol      	&	\texttt{>}\\
	Start terminal symbol       	&	\texttt{"}\\
	End terminal symbol         	&	\texttt{"}\\
	Start option symbol         	&	\texttt{[}\\
	End option symbol           	&	\texttt{]}\\
	Start group symbol          	&	\texttt{(}\\
	End group symbol            	&	\texttt{)}\\
	Start repetition star symbol	&	\texttt{\{}\\
	End repetition star symbol  	&	\texttt{\}}\\
	Start repetition plus symbol	&	\texttt{\{}\\
	End repetition plus symbol		&	\texttt{\}+}\\\hline
\end{tabular}\end{center}

As we know from \cite{BNF} and its research in \cite[\S6.3]{Zaytsev-Thesis2010}, BNF was originally defined as follows:

\begin{center}\begin{tabular}{|l|c|}\hline
	Name	& Value\\\hline
	Defining symbol					&	\backuseq\\
	Definition separator symbol 	&	\backusor\\
	Terminator symbol			 	&	\crleft\\
	Start nonterminal symbol		&	\texttt{<}\\
	End nonterminal symbol      	&	\texttt{>}\\\hline
\end{tabular}\end{center}

While the difference in the appearances of defining symbols is minor and is commonly overlooked, there are several properties of the notation used for
MediaWiki grammar definition that place it well outside BNF, namely:

\begin{itemize}
	\item Using delimiters to explicitly denote terminal symbols
		(instead of using underlined decoration for keywords and relying on implicit assumptions for non-alphanumeric characters).
	\item Presence of comments in the grammar (not in the text around it).
	\item Allowing inconsistent terminator symbol (i.e., a newline or a double newline, sometimes a semicolon).
	\item Having metalanguage symbols for marking optional parts of productions.
	\item Having metalanguage symbols for marking repeated parts of productions.
	\item Having metalanguage symbols for grouping parts of productions.
\end{itemize}

\textbf{Hence, it is not BNF.} For the sake of completeness, let us compare it to the classic EBNF, originally proposed in \cite{EBNF}
(sometimes that dialect is referred to as Wirth Syntax Notation) and standardised much later by ISO as \cite{ISO-EBNF}:

\begin{center}\begin{tabular}{|l|c|c|}\hline
	Name	& Value in WSN	& Value in ISO EBNF\\\hline
	Concatenate symbol				&					&	\texttt{,} \\
	Start comment symbol			&					&	\texttt{(*}\\
	End comment symbol          	&					&	\texttt{*)}\\
	Defining symbol					&	\texttt{=} 		&	\texttt{=} \\
	Definition separator symbol 	&	\texttt{|} 		&	\texttt{|} \\
	Terminator symbol			 	&	\texttt{.} 		&	\texttt{;} \\
	Start terminal symbol       	&	\texttt{"} 		&	\texttt{"} \\
	End terminal symbol         	&	\texttt{"} 		&	\texttt{"} \\
	Start option symbol         	&	\texttt{[} 		&	\texttt{[} \\
	End option symbol           	&	\texttt{]} 		&	\texttt{]} \\
	Start group symbol          	&	\texttt{(} 		&	\texttt{(} \\
	End group symbol            	&	\texttt{)} 		&	\texttt{)} \\
	Start repetition star symbol	&	\texttt{\{}		&	\texttt{\{}\\
	End repetition star symbol  	&	\texttt{\}}		&	\texttt{\}}\\
	Exception symbol	          	&					&	\texttt{-}\\
	Postfix repetition symbol      	&					&	\texttt{*}\\
\hline
\end{tabular}\end{center}

We notice again a list of differences of MediaWiki grammar notation versus WSN and ISO EBNF:

\begin{itemize}
	\item Allowing inconsistent terminator symbol (i.e., a newline or a double newline).
	\item Presence of comments (consistent only with ISO EBNF).
	\item Lack of concatenate metasymbol (consistent only with WSN).
	\item Having metalanguage symbol for exceptions (consistent only with WSN).
	\item Not having a specially designated postfix symbol for denoting repetition (consistent only with WSN).
\end{itemize}

Hence, the notation adopted by MediaWiki grammar, is neither BNF nor EBNF, but an extension of a subset of EBNF.
Since we cannot reuse any previously existing automated grammar extractor, we define this particular notation
with EDD (EBNF Dialect Definition), a part of SLPS (Software Language Processing Suite)~\cite{SLPS} --- and
use \textbf{Grammar Hunter}, a universal configurable grammar extraction tool, for extracting the first
version. The definition itself is a straightforward XML-ification of the first table of this section, so we leave it out of this document.
The only addition is switching on the options of disregarding extra spaces and extra newlines that are left after tokenising the grammar.
The EDD is freely available for re-use in the subversion repository of SLPS\footnote{Available as \linktoedd{config}.}.

\section{Guided grammar extraction}

Since the grammar extraction process is performed for this particular notation for the first time, we use \emph{guided} extraction, when the
results of the extraction are visually compared to the original text by an expert in grammar engineering. This document is a detailed
explanation of observations collected in that process and actions undertaken to resolve the spotted issues.

Given previous experience, it is safe to assume that once the grammar is extracted, we would like to change some parts of it (for grammar
adaptation~\cite{Adaptation}, deyaccification~\cite{DeYacc} and other activities common for grammar recovery~\cite{GRK}). In order for those
changes to stay fully traceable and transparent, we will take the approach of \emph{programmable grammar transformation}. In this methodology,
we take a baseline grammar and an operator suite and by choosing the right operators and parametrising them, we \emph{program} the desired
changes in the same way mainstream programmers use programming languages to create software. These transformation scripts are executable with
the grammar transformation engine: any meta-programming facility would suffice, for this particular work we use XBGF~\cite{XBGF-manual} which
was shown in \cite{JLS-journal} to be the best and the most versatile grammar transformation infrastructure at this moment. The tools of SLPS
that surround XBGF also allow for easy publishing by providing immediate possibilities to transform XBGF scripts to \LaTeX\ or XHTML.

\subsection{Source for extraction}

The grammar of MediaWiki is available on subpages of \cite{MW-BNF}. Striving for more automation, we can use the ``raw'' action to download the
content from the same makefile that performs the extraction\footnote{Available as
\linktofile{Makefile}.}. For example,
the wiki source of Article Title~\cite{MW-BNF-Article-title} is
\url{http://www.mediawiki.org/w/index.php?title=Markup_spec/BNF/Article_title&action=raw}. In order to make our setup stable for the future
when the contents of the wiki page may change (in fact, changing them is one of the main objectives of this work), we can add the revision
number to that command, making it \url{http://www.mediawiki.org/w/index.php?title=Markup_spec/BNF/Article_title&action=raw&oldid=295042}.

\subsection{Article title}\label{article-title}

Parsing Article Title \cite{MW-BNF-Article-title} with Grammar Hunter is not hard and does not report many problems.
One particular peculiarity that we notice when comparing the resulting grammar with the original, is the ``\texttt{...?}'' symbol:

\noindent\begin{boxedminipage}{\textwidth}\footnotesize
\begin{verbatim}
<canonical-page-first-char> ::= <ucase-letter> | <digit> | <underscore> | ...?
<canonical-page-char>       ::= <letter> | <digit> | <underscore> | ...?
\end{verbatim}
\end{boxedminipage}

The ``\texttt{...?}'' symbol is not explained anywhere, but the intuitive meaning is that it is a metasymbol for a possible future extension
point. For example, if in the future one decides to allow a hash symbol (\texttt{\#}) in an article title (currently not allowed for technical
reasons), it will be added as an alternative to the production defining \texttt{canonical-page-char}. The very notion of such extension points
contradicts the contemporary view on language evolution. It is commonly assumed that a grammar engineer cannot predict in advance all the
places in the grammar that will need change in the future: hence, it is better to not mark any of such places explicitly and assume that any
place can be extended, replaced, adapted, transformed, etc. Modern grammar transformation engines such as XBGF~\cite{XBGF-manual},
Rascal~\cite{Rascal} or TXL~\cite{TXL} all have means of extending a grammar in almost any desired place. Since it seems reasonable to remove
these extension points at all, we can do it with XBGF after the extraction\footnote{Part of \linktoxbgf{remove-extension-points}.}:

\noindent
{\footnotesize
\begin{lstlisting}[language=pp]
vertical( in canonical-page-first-char );
removeV(
 canonical-page-first-char:
        "." "." "." "??"
);
horizontal( in canonical-page-first-char );
vertical( in canonical-page-char );
removeV(
 canonical-page-char:
        "." "." "." "??"
);
horizontal( in canonical-page-char );
vertical( in page-first-char );
removeV(
 page-first-char:
        "." "." "." "??"
);
horizontal( in page-first-char );
vertical( in page-char );
removeV(
 page-char:
        "." "." "." "??"
);
horizontal( in page-char );
\end{lstlisting}}

By looking at the grammar where this transformation chain does not apply, one can notice productions in this style:

\noindent\begin{boxedminipage}{\textwidth}\footnotesize
\begin{verbatim}
<canonical-article-title>       ::= <canonical-page> [<canonical-sub-pages>]
<canonical-sub-pages>          ::= <canonical-sub-page> [<canonical-sub-pages>]
<canonical-sub-page>           ::= <sub-page-separator> <canonical-page-chars>
\end{verbatim}
\end{boxedminipage}

In simple words, what we see here is an optional occurrence of a nonterminal called \texttt{canonical-sub-pages}, which is defined as
a list of one or more nonterminals called \texttt{canonical-sub-page}. So, in fact, that optional occurrence consists of zero or more
\texttt{canonical-sub-page} nonterminals. However, these observations are not immediate when looking at the definition, because the
production is written with explicit right recursion. This style of writing productions belong to very early versions of compiler
compilers like YACC~\cite{YACC}, which required manual optimisation of each grammar before parser generation was possible. It has been reported
later on multiple occasions~\cite[etc]{Towards,DeYacc} that it is highly undesirable to perform premature optimisation of a general
purpose grammar for a specific parsing technology that may or may not be used with it at some point in the future. The classic construct
of a list of zero or more nonterminal occurrences is called a Kleene closure~\cite{DragonBook} or Kleene star (since it is commonly denoted
as a postfix star) and is omnipresent in modern grammarware practice.

Using the Kleene star makes the grammars much more concise and readable.
Most parser generators that require right-recursive (or left-recursive) expansions of a Kleene star, can do them automatically on the fly.
Another possible reason for not using a star repetition could have been to stay within limits of pure BNF, but since we have already noted earlier that
this goal was not reached anyway, we see no reason to pretend to seek it. A well-known
grammar beautification technique known as ``deyaccification''~\cite{DeYacc} is performed by the following grammar refactoring
chain\footnote{Part of \linktoxbgf{deyaccify}.}:

\noindent
{\footnotesize
\begin{lstlisting}[language=pp]
massage(
 canonical-sub-pages?,
 (canonical-sub-pages | EPSILON));
distribute( in canonical-sub-pages );
vertical( in canonical-sub-pages );
deyaccify(canonical-sub-pages);
inline(canonical-sub-pages);
massage(
 (canonical-sub-page+ | EPSILON),
 canonical-sub-page*);
massage(
 canonical-page-chars?,
 (canonical-page-chars | EPSILON));
distribute( in canonical-page-chars );
vertical( in canonical-page-chars );
deyaccify(canonical-page-chars);
inline(canonical-page-chars);
massage(
 (canonical-page-char+ | EPSILON),
 canonical-page-char*);
massage(
 sub-pages?,
 (sub-pages | EPSILON));
distribute( in sub-pages );
vertical( in sub-pages );
deyaccify(sub-pages);
inline(sub-pages);
massage(
 (sub-page+ | EPSILON),
 sub-page*);
massage(
 page-chars?,
 (page-chars | EPSILON));
distribute( in page-chars );
vertical( in page-chars );
deyaccify(page-chars);
inline(page-chars);
massage(
 (page-char+ | EPSILON),
 page-char*);
\end{lstlisting}}

Even the simplest metrics can show us that these refactorings have simplified the grammar, reducing it from 15 VAR and 25 PROD to 11 VAR and 17 PROD~\cite{PowerMalloy}, without any fallback in functionality. They have also removed technological idiosyncrasies and improved properties
that are somewhat harder to measure, like readability and understandability.

\subsection{Article}\label{article}

Article~\cite{MW-BNF-Article} contains seven grammar fragments, out of which only the first three conform to the chosen grammar notation.
The last four were copy-pasted from elsewhere and use a different EBNF dialect, which we luckily can also analyse and identify:

\begin{center}\begin{tabular}{|l|c|}\hline
	Name	& Value\\\hline
	Defining symbol					&	\texttt{=}\\
	Definition separator symbol 	&	\texttt{|}\\
	Start special symbol	       	&	\texttt{?}\\
	End special symbol	         	&	\texttt{?}\\
	Start terminal symbol       	&	\texttt{"}\\
	End terminal symbol         	&	\texttt{"}\\
	Start option symbol         	&	\texttt{[}\\
	End option symbol           	&	\texttt{]}\\
	Start group symbol          	&	\texttt{(}\\
	End group symbol            	&	\texttt{)}\\
	Start repetition star symbol	&	\texttt{\{}\\
	End repetition star symbol  	&	\texttt{\}}\\
	Exception symbol 			 	&	\texttt{-}\\
\hline
\end{tabular}\end{center}

We will not lay out its step by step comparison with the notation used in the rest of the MediaWiki grammar, but it suffices to say that the
presence of the exception symbol in the metalanguage is enough to make some grammars inexpressible in a metalanguage without it. BGF does not
have a metasymbol for exception, but we still could express the dialect in EDD\footnote{Available at \linktoedd{metawiki}.} and extract these
parts of the grammar with it. Judging by the presence of the Kleene star in the metalanguage, the grammar engineers who developed those parts
did not intend to stay within BNF limits. Thus, we can also advise to add the use of a plus repetition for denoting a sequence of one or more
nonterminal occurrences, in order to improve readability of productions like these:

\noindent\begin{boxedminipage}{\textwidth}\footnotesize
\begin{verbatim}
Line = PlainText { PlainText } { " " { " " } PlainText { PlainText } } ;
Text = Line { Line } { NewLine { NewLine } Line { Line } } ;
\end{verbatim}
\end{boxedminipage}

Or, in postfix-oriented BNF that we use within SLPS:

\noindent\begin{boxedminipage}{\textwidth}\footnotesize
\begin{lstlisting}[language=pp]
Line:
        PlainText PlainText* (" " " "* PlainText PlainText*)*
Text:
        Line Line* (NewLine NewLine* Line Line*)*
\end{lstlisting}
\end{boxedminipage}

Compare with the version that we claim to be more readable:

\noindent\begin{boxedminipage}{\textwidth}\footnotesize
\begin{lstlisting}[language=pp]
Line:
        PlainText+ (" "+ PlainText+)*
Text:
        Line+ (NewLine+ Line+)*
\end{lstlisting}
\end{boxedminipage}

In fact, many modern grammar definition formalisms have a metaconstruct called ``separator list'', because \texttt{Text} above is nothing more
than a (multiple) \texttt{Newline}-separated list of \texttt{Line}s. We do not enforce this kind of metaconstructs here,
but we do emphasize the fact that the very understanding of \texttt{Text} being a separated list of \texttt{Line}s
was not clear before our proposed refactoring. In the case if MediaWiki still wants the grammar representation to have only one type of
repetition or even no repetition at all, such a view can be automatically derived from the baseline grammar preserved in a more expressive
metalanguage. 
The refactorings that utilise the plus notation are rather straightforward\footnote{Part of \linktoxbgf{utilise-repetition}.}:

\noindent
{\footnotesize
\begin{lstlisting}[language=pp]
massage(
 PlainText PlainText*,
 PlainText+);
massage(
 Line Line*,
 Line+);
massage(
 NewLine NewLine*,
 NewLine+);
massage(
 " " " "*,
 " "+);
\end{lstlisting}}

Further investigation draws our attention to these productions:

\noindent\begin{boxedminipage}{\textwidth}\footnotesize
\begin{verbatim}
PageName = TitleCharacter , { [ " " ] TitleCharacter } ;
PageNameLink = TitleCharacter , { [ " " | "_" ] TitleCharacter } ;
\end{verbatim}
\end{boxedminipage}

The comma used in both productions is not a terminal symbol ``\texttt{,}'': in fact, it is a concatenate symbol from ISO EBNF~\cite{ISO-EBNF}. Since ISO EBNF
is not the notation used, the commas must have been left out unintentionally---this is what usually happens when grammars are transformed
manually and not in a disciplined way. Grammar Hunter assumed that the quotes were forgotten in this place (since a comma is not a good name
for a nonterminal), so we need to project it away (the corresponding operator is called \textbf{abstractize} because it shifts
a grammar from concrete syntax to abstract syntax). These are the transformations that we write down\footnote{Complete listing of \linktoxbgf{remove-concatenation}.}:

\noindent
{\footnotesize
\begin{lstlisting}[language=pp]
abstractize(
 PageName:
        TitleCharacter <","> (" "? TitleCharacter)*
);
abstractize(
 PageNameLink:
        TitleCharacter <","> ((" " | "_")? TitleCharacter)*
);
\end{lstlisting}}

The following fragment uses excessive bracketing: parenthesis are used to group symbols together, which is usually necessary for inner choices
and similar cases when one needs to override natural priorities. However, in this case it is unnecessary:

\noindent\begin{boxedminipage}{\textwidth}\footnotesize
\begin{verbatim}
SectionTitle = ( SectionLinkCharacter - "=" )
                                    { [ " " ] ( SectionLinkCharacter - "=" ) } ;
LinkTitle = { UnicodeCharacter { " " } } ( UnicodeCharacter - "]" ) ;
\end{verbatim}
\end{boxedminipage}

Excessive bracketing is not a problem for SLPS toolset since all BGF grammars are normalised before serialisation, and it includes a step
of refactoring trivial subsequences, but we still report it for the sake of reproducibility within a different environment.

The following grammar production uses a strange-looking construction that is explained in the text to be the ``non-greedy'' variant of the optional newline:

\noindent\begin{boxedminipage}{\textwidth}\footnotesize
\begin{verbatim}
<special-block-and-more>  ::=
                 <special-block> ( EOF | [<newline>] <special-block-and-more> 
                                       | (<newline> | "") <paragraph-and-more> )
\end{verbatim}
\end{boxedminipage}

The purpose of a syntax definition such as a BNF is to define syntax of a language. Thus, any references to the semantics of the parsing process
should be avoided. The definition of ``greediness'' as ordered alternatives, given at the first page of \cite{MW-BNF}, contradicts the classic
definition based on token consumption, and contradicts the basics of EBNF. Approaches alternative to context-free grammars such as PEG~\cite{PEG}
should be considered if ordered alternatives are really required. For EBNF (or BGF), we refactor the singularity as
follows\footnote{Part of \linktoxbgf{utilise-question}.}:

\noindent
{\footnotesize
\begin{lstlisting}[language=pp]
massage(
 (newline | EPSILON),
 newline?);
\end{lstlisting}}

Since at this point the subgrammar of this part must be rather consistent, we can execute some simple grammar analyses to help assess the grammar quality.
One of them is based on a well-known notion of bottom and top nonterminals~\cite{DeYacc,PLEX}: a top is one that is defined but never used;
a bottom is one that is used but never defined. We were surprised to see \texttt{WhiteSpaces} in the list of top nonterminals, while
\texttt{Whitespaces} was in the list of bottom nonterminals. Apparently, a renaming is needed\footnote{Part of \linktoxbgf{unify-whitespace}.}:

\noindent
{\footnotesize
\begin{lstlisting}[language=pp]
unite(WhiteSpaces, Whitespaces);
\end{lstlisting}}

The definition of nonterminal \texttt{BlockHTML} contains textual annotation claiming that it is not yet referred to. We decided to parse it
anyway and validate that assertion afterwards. Indeed, it showed up as an unconnected grammar fragment, which we can then safely
remove\footnote{Part of \linktoxbgf{connect-grammar}.}:

\noindent
{\footnotesize
\begin{lstlisting}[language=pp]
eliminate(BlockHTML);
\end{lstlisting}}

\subsection{Noparse block}\label{noparse-block}

Apart from quoting the language name in the source tags, which makes the start grammar symbol change from \texttt{<source lang=bnf>}
to \texttt{<source lang="bnf"{}>}, the Noparse Block~\cite{MW-BNF-Noparse-block} uses the same EBNF dialect that we derived as the
starting step of our extraction. However, there are two major exceptions:

\begin{itemize}
	\item Round brackets and square brackets have swapped their meaning.
	\item A lookahead assertion metasymbol is used, borrowed from Perl Compatible Regular Expressions library.
\end{itemize}

The first impression given by cursory examination of the extracted grammar is that it uses excessive bracketing (mentioned in the previous section):

\noindent\begin{boxedminipage}{\textwidth}\footnotesize
\begin{verbatim}
<pre-block>               ::= <pre-opening-tag> (<whitespace>) <pre-body>
                                   (<whitespace>) [<pre-closing-tag> | (?=EOF) ]
<pre-opening-tag>         ::= "&lt;pre" (<whitespace> (<characters>)) "&gt;"
<pre-closing-tag>         ::= "&lt;/pre" (<whitespace>) "&gt;"
<pre-body>                ::= <characters>
\end{verbatim}
\end{boxedminipage}

However, if we assume this to be true, the meaning of the grammar will become inadequate: for example, it will have mandatory whitespace in many places.
On the other hand, making the last part of the grammar (\texttt{<nowiki-closing-tag> | (?=EOF)}) optional is also inadequate,
because optional assertion will never make sense. This particular lookahead assertion is displayed as \texttt{(?=EOF)} and means
basically an $\varepsilon$ that must be followed by EOF (even that definition is not that
apparent from the low-level description saying \emph{``It asserts that an EOF follows, but does not consume the EOF.''}).
The presence or absence of lookahead based facilities is heavily dependent on the parsing technology, and therefore should be avoided as much as
possible, as noted by multiple sources~\cite{Towards,Recovery,DeYacc}.
More straightforward and high level assertions like ``should be followed by'' and ``should not be followed by''
are available in modern metaprogramming languages like Rascal~\cite{Rascal} instead.

Since the general problem of leaving opened tags at the end of the article text is much bigger than the tags described in this part of the grammar,
we opt for removing these assertions altogether and solving the problem later with suitable technology. EBNF has never been intended for and has never
been good at defining tolerant parsers~\cite{Tolerant}. Since we have to construct another EBNF dialect in order to parse the Noparse Block fragment
correctly anyway, we specify ``\texttt{(?=EOF)}'' as a notation for $\varepsilon$ (otherwise we would have to fix the problem later with
a horizontal remove operator from XBGF). Those explicit empty sequence metasymbols need to be refactored into proper optional
symbols\footnote{Part of \linktoxbgf{remove-lookahead}.}:

\noindent
{\footnotesize
\begin{lstlisting}[language=pp]
massage(
 (nowiki-closing-tag | EPSILON),
 nowiki-closing-tag?);
massage(
 (pre-closing-tag | EPSILON),
 pre-closing-tag?);
massage(
 (html-closing-tag | EPSILON),
 html-closing-tag?);
\end{lstlisting}}

In every notation that comprises similar looking symbols and metasymbols that can be encountered within the same context, there is need for
\emph{escaping} some special characters. In this part of the MediaWiki grammar escaping is done in HTML entities, which is not explainable
with grammar-based arguments. However, we recall that our extraction source is a handcrafted grammar that was meant to reproduce the behaviour
of the MediaWiki PHP parse---so, in a sense, it was (manually) extracted, and what we have just encountered is in fact a legacy artefact randomly
inherited from its source. Such legacy should be removed by following transformation steps\footnote{Complete listing of \linktoxbgf{dehtmlify}.}:

\noindent
{\footnotesize
\begin{lstlisting}[language=pp]
renameT("&lt;nowiki", "<<nowiki");
renameT("&lt;/nowiki", "<</nowiki");
renameT("&lt;pre", "<<pre");
renameT("&lt;/pre", "<</pre");
renameT("&lt;html", "<<html");
renameT("&lt;/html", "<</html");
renameT("&lt;!--", "<<!--");
replace("&gt;", ">>");
\end{lstlisting}}

There are two more problems in the Noparse Block part that concern the nonterminal \texttt{characters}. First, it is undefined (bottom). As we will see in
\S\ref{inline}, there is a nonterminal called \texttt{character}---issues like these with ``forgetting'' to define some nonterminals
with readable names are quite common in handcrafted grammars, as noted by \cite{NeedsGrammarware2005} and other sources. A trivially
guessed definition for \texttt{characters} is either ``one-or-more'' or ``zero-or-more'' repetition of \texttt{character}. Since \texttt{characters}
is mostly used as an optional nonterminal, we assume that it is one or more\footnote{Part of \linktoxbgf{connect-grammar}.}:

\noindent
{\footnotesize
\begin{lstlisting}[language=pp]
define(
 characters:
        character+
);
\end{lstlisting}}

The second problem is its usage in \texttt{html-comment} (remember that round brackets mean optionality here):

\noindent\begin{boxedminipage}{\textwidth}\footnotesize
\begin{verbatim}
<html-comment>             ::= "&lt;!--" ({ characters }) "-->"
\end{verbatim}
\end{boxedminipage}

Since we do not need to make a Kleene repetition optional, we can refactor it as follows\footnote{Part of \linktoxbgf{refactor-repetition}.}:

\noindent
{\footnotesize
\begin{lstlisting}[language=pp]
unfold(characters in html-comment);
massage(
 character+*,
 character*);
massage(
 character*?,
 character*);
massage(
 character*,
 character+?);
fold(characters in html-comment);
\end{lstlisting}}

More detailed information about leaving combinations of various kinds of repetition and optionality in the deployed grammar will be
given in the next section.

\subsection{Links}\label{links}

\begin{figure}
	\centerline{\includegraphics[width=0.8\textwidth]{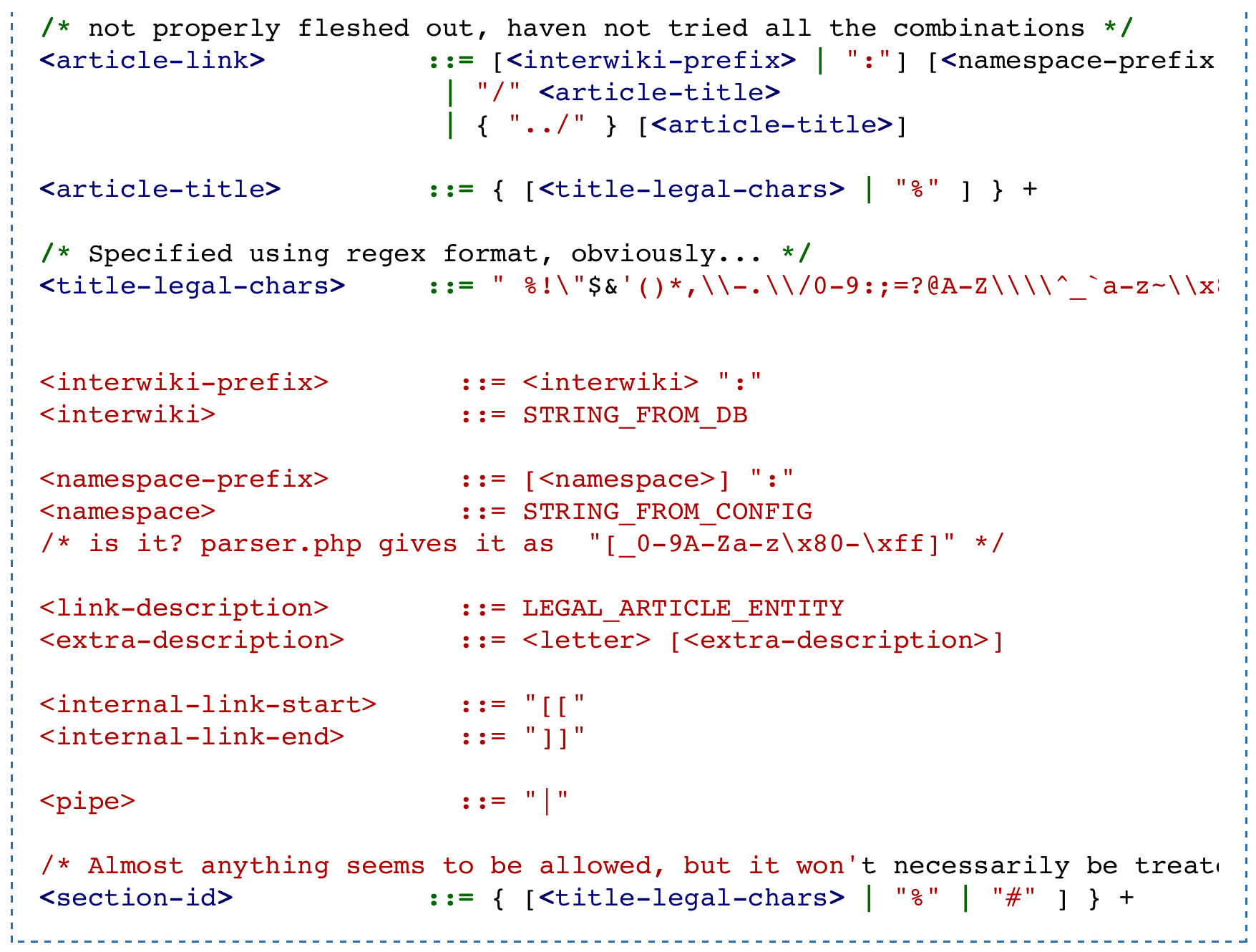}}
	\caption{A syntax that even MediaWiki cannot colour-code properly~\cite{MW-BNF-Links}.}
	\label{F:mediawiki}
\end{figure}

Links definitions~\cite{MW-BNF-Links} exhibit bits of yet another notation, namely the one where a set of possible values is given, assuming that
only one should be picked. In the MediaWiki grammar it is erroneously called a ``regex format''---regular expressions do use this notation
in some places, but not everywhere and it is not exclusive to them. This notation is very much akin to ``one-of'' metaconstructs also encountered
in definitions of other software languages such as C\#~\cite[\S3.2.4]{Zaytsev-Thesis2010}. In the MediaWiki grammar, it looks like this:

\noindent\begin{boxedminipage}{\textwidth}\footnotesize
\begin{verbatim}
/* Specified using regex format, obviously... */
<title-legal-chars>  ::= " %!\"$&'()*,\\-.\\/0-9:;=?@A-Z\\\\^_`a-z~\\x80-\\xFF+"
\end{verbatim}
\end{boxedminipage}

The unobviousness of the notation is perfectly simplified by the fact that even the MediaWiki engine itself fails to parse and colour-code
it correctly, as seen on \autoref{F:mediawiki}. In fact, when we look at the expression more closely, we can notice that it is even
incorrect in itself, since it uses double-escaping for most backslashes (ruining them) and does not escape the dot (which denotes any character
when unescaped). Some other characters like \texttt{*} or \texttt{+} should arguably also be escaped, but it is impossible to decide firmly
on escaping rules when we have no engine to process this string. However, the correct expression should have looked similar to this:

\noindent\begin{boxedminipage}{\textwidth}\footnotesize
\begin{verbatim}
<title-legal-chars>  ::= " %!\"$&'()*,\-\.\/0-9:;=?@A-Z\\^_`a-z~\x80-\xFF+"
\end{verbatim}
\end{boxedminipage}

\smallskip Which we rewrite as (some invisible characters are omitted for readability):\smallskip 

\noindent\begin{boxedminipage}{\textwidth}\footnotesize
\centerline{\includegraphics[width=\textwidth]{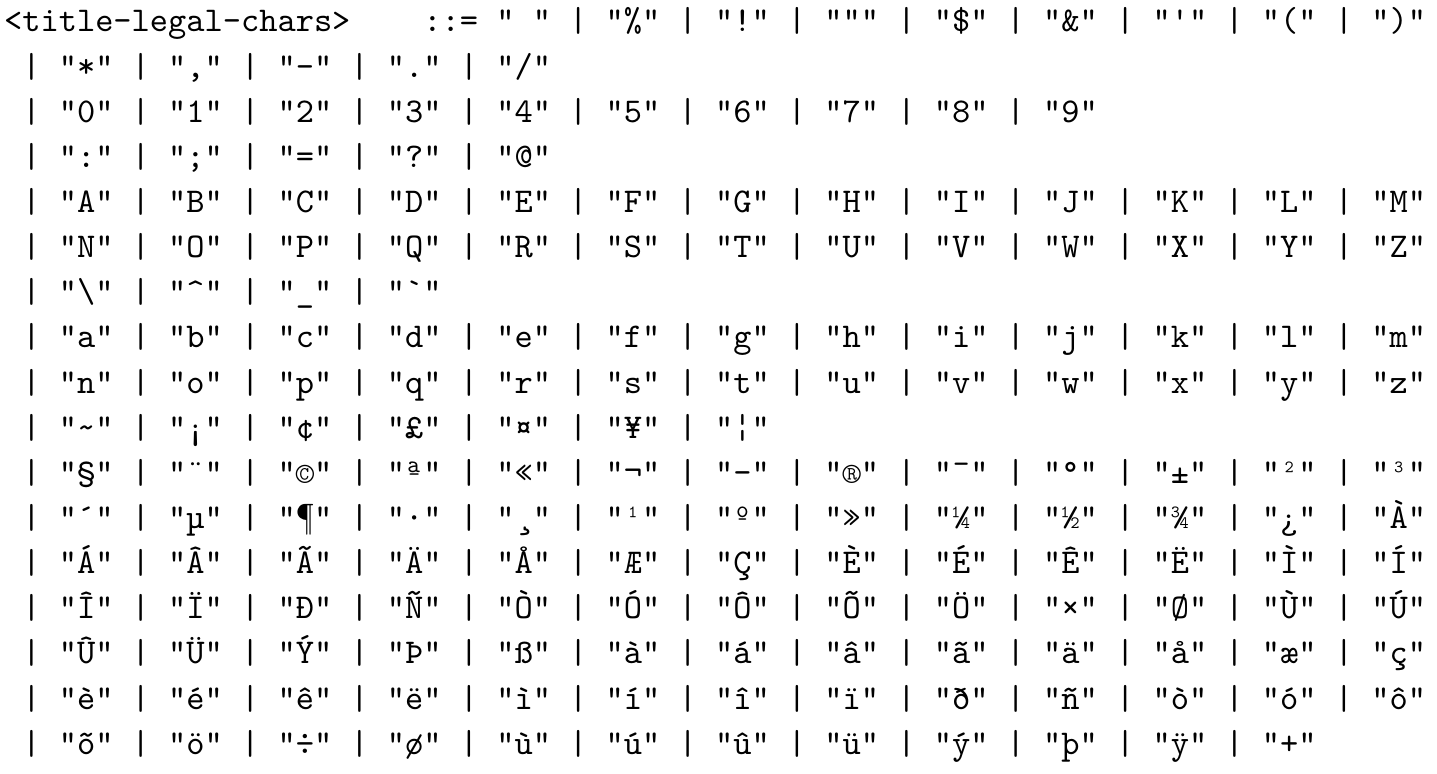}}
\end{boxedminipage}

This refactored version with all alternatives given explicitly was created automatically by a trivial Python one-liner and can be parsed
without any trouble by Grammar Hunter. We should also note that the name for this nonterminal is misleading, since it represents only one
character. This is not a technical mistake, but we can improve \emph{learnability} of the grammar by fixing it\footnote{Part of \linktoxbgf{fix-names}.}:

\noindent
{\footnotesize
\begin{lstlisting}[language=pp]
renameN(title-legal-chars, title-legal-char);
\end{lstlisting}}

Grammar Hunter displays an error message but is capable of dealing with this fragment:

\noindent\begin{boxedminipage}{\textwidth}\footnotesize
\begin{verbatim}
<article-link> ::=
                 [<interwiki-prefix> | ":" ] [<namespace-prefix] <article-title>
\end{verbatim}
\end{boxedminipage}

The problem in this grammar production is in ``\texttt{[<namespace-prefix]}'' (note the unbalanced angle brackets).
The start nonterminal symbol here is followed by
the name of the nonterminal and then by the end option symbol without the end nonterminal symbol. This kind of problems are rather
common in grammars that have been created manually and have never been tested in any environment that would make them executable or
validate consistency otherwise. Grammar Hunter can resolve this problem by using the heuristic of next best guess, which is to assume
that the nonterminal name ended at the first alphanumeric/non-alphanumeric border that happened after the unbalanced start nonterminal symbol.

Next, consider the following two grammar productions that lead to several problems simultaneously:

\noindent\begin{boxedminipage}{\textwidth}\footnotesize
\begin{verbatim}
<article-title>         ::= { [<title-legal-chars> | "%" ] } +
<section-id>            ::= { [<title-legal-chars> | "%" | "#" ] } +
\end{verbatim}
\end{boxedminipage}

As we have explained above, the grammar notation used for the MediaWiki grammar was never defined explicitly in any formal or informal way,
so we had to infer it in \S\ref{ceci-n'est-pas-une-bnf}. When inferring its semantics, we had two options: to treat the plus as a postfix
metasymbol or to treat ``\texttt{\{}'' and ``\texttt{\}+}'' as bracket metasymbols. Both variants are possible and feasible, since
Grammar Hunter is capable of dealing with ambiguous starting metasymbols (i.e., ``\texttt{\{}'' as both a start repetition star symbol and
a start repetition plus symbol). We obviously opt for the latter variant because from the formal language theory we all know that for any
$x$ it is always true that $(x^*)^+\equiv x^*$, so a postfix plus operation on a star repetition is useless and we tend to
\href{http://en.wikipedia.org/wiki/Wikipedia:Assume_good_faith}{assume good faith} of grammar engineers who made use of it.
But even if we assume it to be a transitive closure (a plus repetition), which is a common notation for a sequence of one or more occurrences of
a subexpression, the productions become parseable, but they are bound to deliver problems with ambiguities~\cite{Basten2010}
on later stages of grammar deployment, since in these particular grammar fragments optional symbols are iterated.

To give a simple example, suppose we have a nonterminal $x$ defined as $a^+$, and $a$ itself is defined as $``a"?$ (either $``a"$ or $\varepsilon$).
Then the following are two distinct possibilities to parse $``aa"$ with such a grammar:

\[
\begin{array}{ccccccc}
a^+&&&&        & a^+ &\\
\nnearrow\:\:\:\nnwarrow&&&&        \phantom{xx}\nearrow&\uparrow&\nwarrow\phantom{xx}\\
a\qquad a&&&&        a&a&a\\
\uparrow\qquad\uparrow&&&&        \uparrow&\uparrow&\uparrow\\
``a"\quad ``a"&&&&        ``a"&\varepsilon&``a"\\
\end{array}
\]

The number of such ways to parse even the simplest of expressions is infinite, and special algorithms need to be utilised to detect
such problems at the parser generator level. Thus, to prevent this trouble from happening, we massage the productions above to use a simple
star repetition instead, which is an equivalent unambiguous construct\footnote{Part of \linktoxbgf{utilise-repetition}.}:

\noindent
{\footnotesize
\begin{lstlisting}[language=pp]
massage(
 (title-legal-chars | "%")?+,
 (title-legal-chars | "%")*);
massage(
 (title-legal-chars | "%" | "#")?+,
 (title-legal-chars | "%" | "#")*);
\end{lstlisting}}

Reading further, we notice one of the nonterminals being defined with explicit right recursion:

\noindent\begin{boxedminipage}{\textwidth}\footnotesize
\begin{verbatim}
<extra-description>       ::= <letter> [<extra-description>]
\end{verbatim}
\end{boxedminipage}

The problem is known and has been discussed above, all we need here is proper deyaccification\footnote{Part of \linktoxbgf{deyaccify}.}:

\noindent
{\footnotesize
\begin{lstlisting}[language=pp]
massage(
 extra-description?,
 (extra-description | EPSILON)
 in extra-description);
distribute( in extra-description );
vertical( in extra-description );
deyaccify(extra-description);
\end{lstlisting}}

The last problem with the Links part of the grammar is the use of natural language inside a BNF production:

\noindent\begin{boxedminipage}{\textwidth}\footnotesize
\begin{verbatim}
<protocol>          ::= ALLOWED_PROTOCOL_FROM_CONFIG (e.g. "http://", "mailto:")
\end{verbatim}
\end{boxedminipage}

Examples are never a part of a syntax definition: the alternatives are either listed exhaustively
(like we will do later when we make the grammar complete)
or belong in the comments (like it was undoubtedly intended here).
A projection is needed to remove them from the raw extracted grammar\footnote{Complete listing of \linktoxbgf{remove-comments}.}:

\noindent
{\footnotesize
\begin{lstlisting}[language=pp]
project(
 protocol:
        ALLOWED_PROTOCOL_FROM_CONFIG <(e "." g "." "http://" "," "mailto:")>
);
\end{lstlisting}}

\subsection{Magic links}\label{magic-links}

Just like Noparse Block discussed above in \S\ref{noparse-block}, Magic Links~\cite{MW-BNF-Magic-links} also uses \texttt{<source lang="bnf"{}>}
as the start grammar symbol, but this is the least problem encountered in this fragment. Consider the following productions:

\noindent\begin{boxedminipage}{\textwidth}\footnotesize
\begin{verbatim}
<isbn>          ::= "ISBN" (" "+) <isbn-number> ?(non-word-character /\b/)
<isbn-number>  ::= ("97" ("8" | "9") (" " | "-")?) (DIGIT (" " | "-")?)
                                                        {9} (DIGIT | "X" | "x")
\end{verbatim}
\end{boxedminipage}

We see a notation where:

\begin{itemize}
	\item A postfix plus repetition metasymbol is used, which is not encountered anywhere else in the MediaWiki.
	\item The character used as the postfix repetition metasymbol clashes with end repetition plus metasymbol from
		Inline Text~\cite{MW-BNF-Inline-text} and Links~\cite{MW-BNF-Links}\footnote{Indirect clash of ``\texttt{\}+}'' being an end repetition plus symbol as well as a sequence of an end repetition star symbol and a postfix repetition metasymbol.}.
	\item A postfix optionality metasymbol is used, which is not encountered anywhere else in the MediaWiki.
	\item The character used as the postfix optionality metasymbol clashes with start special metasymbol and end special metasymbol
		from Article~\cite{MW-BNF-Article}, Inline Text~\cite{MW-BNF-Inline-text} and Special Block~\cite{MW-BNF-Special-block}.
	\item The same character used as the postfix optionality metasymbol is used as in a prefix notation that relies on lookahead.
	\item A regular expression is used inside the lookahead assertion.
	\item A terminal symbol (``9'') is not explicitly marked as such.
	\item A nonterminal symbol (``DIGIT'') is not explicitly marked as such.
\end{itemize}

Along with the discussion from \S\ref{noparse-block}, we first remove the lookahead assertions. They (arguably) do not belong in
EBNF at all, and definitely do not belong in such a form\footnote{Part of \linktoxbgf{remove-lookahead}.}:

\noindent
{\footnotesize
\begin{lstlisting}[language=pp]
project(
 isbn:
        "ISBN" " " "+" isbn-number <("?" non-word-character "/" "\" b "/")>
);
\end{lstlisting}}

We do not even try to add the postfix plus repetition metasymbol to the notation definition, since it is used only once, since it
clashes with something else, and since there is a special nonterminal \texttt{spaces} that should be used instead
anyway\footnote{Part of \linktoxbgf{unify-whitespace}.}:

\noindent
{\footnotesize
\begin{lstlisting}[language=pp]
replace(
 " " "+",
 spaces);
\end{lstlisting}}

Then we adjust the grammar for the untreated postfix question metasymbol\footnote{Part of \linktoxbgf{utilise-question}.}:

\noindent
{\footnotesize
\begin{lstlisting}[language=pp]
abstractize(
 isbn-number:
        "97" ("8" | "9") (" " | "-") <"??"> DIGIT (" " | "-") <"??"> "9"*
                (DIGIT | "X" | "x")
);
widen(
 (" " | "-"),
 (" " | "-")?
 in isbn-number);
\end{lstlisting}}

\subsection{Special block}\label{special-block}

Just as in \cite{MW-BNF-Article-title}, the Special Block uses a special metasymbol for omitted grammar fragments~\cite{MW-BNF-Special-block}.
This case is subtly different from the one discussed in \S\ref{article-title} in a sense that it explicitly says in the accompanying text that
``The dots need to be filled in''. This information is undoubtedly useful, but considering the fact that its very presence renders the grammar
non-executable, we decide to remove it from the grammar and let the documentation tell the story about how much of the intended language
does the grammar cover\footnote{Part of \linktoxbgf{remove-extension-points}.}:

\noindent
{\footnotesize
\begin{lstlisting}[language=pp]
vertical( in special-block );
removeV(
 special-block:
        "." "." "."
);
horizontal( in special-block );
\end{lstlisting}}

In the same first production there is an alternative that reads \texttt{<nowiki><table></nowiki>}, which seems like either a leftover
after manually cleaning up the markup, or a legacy escaping trick. Either way, \texttt{nowiki} wrapping is not necessary for displaying
this fragment and is generally misleading: the chevrons around ``\texttt{table}'' mean to denote it explicitly as a nonterminal, not as an HTML tag.
We project away the unnecessary parts\footnote{Part of \linktoxbgf{fix-markup}.}:

\noindent
{\footnotesize
\begin{lstlisting}[language=pp]
vertical( in special-block );
project(
 special-block:
        <nowiki> table </ nowiki>
);
horizontal( in special-block );
\end{lstlisting}}

There are also more cases of excessive bracketing which are fixed automatically by Grammar Hunter:

\noindent\begin{boxedminipage}{\textwidth}\footnotesize
\begin{verbatim}
<defined-term>            ::= ";" <text> [ (<definition>)]
\end{verbatim}
\end{boxedminipage}

A nonterminal symbol called \texttt{dashes} is arguably superfluous and can be replaced by a Kleene star of a dash terminal:

\noindent\begin{boxedminipage}{\textwidth}\footnotesize
\begin{verbatim}
<horizontal-rule>         ::= "----" [<dashes>] [<inline-text>] <newline>
<dashes>                  ::= "-" [<dashes>]
\end{verbatim}
\end{boxedminipage}

Still, we can keep it in the grammar for the sake of possible future BNF-ification, but refactor the idiosyncrasy
(the right recursion)\footnote{Part of \linktoxbgf{deyaccify}.}:

\noindent
{\footnotesize
\begin{lstlisting}[language=pp]
massage(
 dashes?,
 (dashes | EPSILON)
 in dashes);
distribute( in dashes );
vertical( in dashes );
deyaccify(dashes);
\end{lstlisting}}

The worst part of the Special Block part is the section titled ``Tables'': it contains eight productions in a different notation, with
a comment ``From meta...minor reformatting''. This reformatting has obviously been performed manually, since it does not utilise
the standard notation of the rest of the grammar, nor is it compatible with the MetaWiki notation that we have encountered in \S\ref{article}:
the defining symbol is from the MediaWiki notation, the terminator symbol is from the MetaWiki notation, etc:

\begin{center}\begin{tabular}{|l|c|}\hline
	Name	& Value\\\hline
	Defining symbol					&	\texttt{::=}\\
	Terminator symbol				&	\texttt{;}\\
	Definition separator symbol 	&	\texttt{|}\\
	Start special symbol	       	&	\texttt{?}\\
	End special symbol	         	&	\texttt{?}\\
	Start terminal symbol       	&	\texttt{"}\\
	End terminal symbol         	&	\texttt{"}\\
	Start nonterminal symbol       	&	\texttt{<}\\
	End nonterminal symbol         	&	\texttt{>}\\
	Start option symbol         	&	\texttt{[}\\
	End option symbol           	&	\texttt{]}\\
\hline
\end{tabular}\end{center}

To save the trouble of post-extraction fixing, we used this configuration as a yet another EDD file to extract this grammar fragment and merge it
with the rest of the grammar. The naming convention of the fragment is still not synchronised with the rest (i.e., camel case vs.\ dash-separated lowercase),
but we will deal with it later in \S\ref{beautification}.

We also see a problem similar to the one discussed above in \S\ref{noparse-block}, namely an optional zero-or-more repetition:

\noindent\begin{boxedminipage}{\textwidth}\footnotesize
\begin{verbatim}
<space-block>             ::= " " <inline-text> <newline> [ {<space-block-2} ]
\end{verbatim}
\end{boxedminipage}

The solution is also already known to us\footnote{Part of \linktoxbgf{refactor-repetition}.}:

\noindent
{\footnotesize
\begin{lstlisting}[language=pp]
massage(
 space-block-2*?,
 space-block-2*);
\end{lstlisting}}

When comparing the list of top nonterminals with the list of bottom ones, we notice \texttt{TableCellParameters} being used while
\texttt{TableCellParameter} being defined. Judging by its clone named \texttt{TableParameters}, the intention was to name it plural, so
we perform unification\footnote{Part of \linktoxbgf{fix-names}.}:

\noindent
{\footnotesize
\begin{lstlisting}[language=pp]
unite(TableCellParameter, TableCellParameters);
\end{lstlisting}}

\subsection{Inline text}\label{inline}

Suddenly, \cite{MW-BNF-Inline-text} uses bulleted-list notation for listing alternatives in a grammar:

\noindent\begin{boxedminipage}{\textwidth}\footnotesize
\begin{verbatim}
<text-with-formatting>    ::=
                             | <formatting> 
                             | <inline-html> 
                             | <noparseblock>
                             | <behaviour-switch> 
                             | <open-guillemet> | <close-guillemet>
                             | <html-entity>
                             | <html-unsafe-symbol>
                             | <text>
                             | <random-character> 
                             | (more missing?)...
\end{verbatim}
\end{boxedminipage}

This is almost never encountered in grammar engineering, but not completely unknown to computer science---for example, TLA$^+$ uses this
notation~\cite{TLAplus}.
In our case it is confusing for Grammar Hunter since newlines are also used in the notation to separate production rules, and since
it only happens in two productions, we decide to manually remove the first bar there. The last line of the sample above also shows an extension
point discussed earlier in \S\ref{article-title} and \S\ref{special-block}, which we remove\footnote{Part of \linktoxbgf{remove-extension-points}.}:

\noindent
{\footnotesize
\begin{lstlisting}[language=pp]
vertical( in text-with-formatting );
removeV(
 text-with-formatting:
        (more missing "??") "." "." "."
);
horizontal( in text-with-formatting );
\end{lstlisting}}

Nonterminal \texttt{noparseblock} is referenced in the same grammar fragment, but never encountered elsewhere in the grammar,
later we will unite it with \texttt{noparse-block} when specifically considering enforcing consistent naming convention in
\S\ref{naming-convention}.

The next problematic fragment is the following:

\noindent\begin{boxedminipage}{\textwidth}\footnotesize
\begin{verbatim}
<html-entity-name>      ::= Sanitizer::$wgHtmlEntities (case sensitive)
 (* "Aacute" | "aacute" | ... *)
\end{verbatim}
\end{boxedminipage}

It has three problems:

\begin{itemize}
	\item Referencing PHP variables from the grammar is unheard of.
	\item Static semantics within postfix parenthesis in plain English is not helpful.
	\item A comment that uses ``\texttt{(*}'' and ``\texttt{*)}'' as delimiters instead of
		``\texttt{/*}'' and ``\texttt{*/}'' used in the rest of the grammar.
\end{itemize}

These identified problems can be solved with projecting excessive symbols, leaving only one nonterminal reference,
which will remain undefined for now\footnote{Complete listing of \linktoxbgf{remove-php-legacy}.}:

\noindent
{\footnotesize
\begin{lstlisting}[language=pp]
project(
 html-entity-name:
        <(Sanitizer ":" ":" "$")> wgHtmlEntities <(case sensitive (("*" "Aacute")
                        | "aacute" | ("." "." "." "*")))>
);
\end{lstlisting}}

Later in \S\ref{wgHtmlEntities} we will reuse the source code of \texttt{Sanitizer} class to formally complete the grammar
by defining \texttt{wgHtmlEntities} nonterminal.

The following fragment combines two double problems that have already been encountered before. The first problem is akin to the one we
have noticed in \S\ref{links}, namely having a nonterminal with ``\texttt{-characters}'' in its name, which is supposed to denote
only one character taken from a character class; the second part of that problem is the usage of the regular expression notation.
The second problem is an omission/extension point (cf.\ \S\ref{article-title} and \S\ref{special-block}), which is expressed in Latin:

\noindent\begin{boxedminipage}{\textwidth}\footnotesize
\begin{verbatim}
<harmless-characters>    ::= /[A-Za-z0-9] etc
\end{verbatim}
\end{boxedminipage}

We rewrite it as follows:\smallskip

\noindent\begin{boxedminipage}{\textwidth}\footnotesize
\begin{verbatim}
<harmless-characters>    ::=
   "A" | "B" | "C" | "D" | "E" | "F" | "G" | "H" | "I" | "J" | "K" | "L" | "M"
 | "N" | "O" | "P" | "Q" | "R" | "S" | "T" | "U" | "V" | "W" | "X" | "Y" | "Z"
 | "a" | "b" | "c" | "d" | "e" | "f" | "g" | "h" | "i" | "j" | "k" | "l" | "m"
 | "n" | "o" | "p" | "q" | "r" | "s" | "t" | "u" | "v" | "w" | "x" | "y" | "z"
 | "0" | "1" | "2" | "3" | "4" | "5" | "6" | "7" | "8" | "9"
\end{verbatim}
\end{boxedminipage}

The name of the nonterminal symbol \texttt{harmless-characters} is misleading, since it represents only one
character. In fact, simple investigation into top and bottom nonterminals~\cite{Recovery} shows that it
is not referenced anywhere in the grammar, but a nonterminal \texttt{harmless-character} is used in the definition
of \texttt{text}. Hence, we want to unite those two nonterminals\footnote{Part of \linktoxbgf{fix-names}.}:

\noindent
{\footnotesize
\begin{lstlisting}[language=pp]
unite(harmless-characters, harmless-character);
\end{lstlisting}}

The immediately following production contains a special symbol written in the style of ISO EBNF and MetaWiki:

\noindent\begin{boxedminipage}{\textwidth}\footnotesize
\begin{verbatim}
<random-character>       ::= ? any character ... ?
\end{verbatim}
\end{boxedminipage}

Instead of adjusting the assumed notation definition, we choose to let Grammar Hunter parse it as it is, and to subsequently
transform the result to a special BGF metasymbol with the same semantics (i.e., ``any character'')\footnote{Part of \linktoxbgf{define-lexicals}.}:

\noindent
{\footnotesize
\begin{lstlisting}[language=pp]
redefine(
 random-character:
        ANY
);
\end{lstlisting}}

The next problematic fragment once again contains omission/extension points:

\noindent\begin{boxedminipage}{\textwidth}\footnotesize
\begin{verbatim}
<ucase-letter>          ::= "A" | "B" | ... | "Y" | "Z"
<lcase-letter>          ::= "a" | "b" | ... | "y" | "z"
<decimal-digit>         ::= "0" | "1" | ... | "8" | "9"
\end{verbatim}
\end{boxedminipage}

Since in fact they represent all possible alternatives from the given range,
we rewrite them as follows:\smallskip

\noindent\begin{boxedminipage}{\textwidth}\footnotesize
\begin{verbatim}
<ucase-letter>          ::=
  "A" | "B" | "C" | "D" | "E" | "F" | "G" | "H" | "I" | "J" | "K" | "L" | "M"
| "N" | "O" | "P" | "Q" | "R" | "S" | "T" | "U" | "V" | "W" | "X" | "Y" | "Z"
<lcase-letter>          ::=
   "a" | "b" | "c" | "d" | "e" | "f" | "g" | "h" | "i" | "j" | "k" | "l" | "m"
 | "n" | "o" | "p" | "q" | "r" | "s" | "t" | "u" | "v" | "w" | "x" | "y" | "z"
<decimal-digit>         ::=
   "0" | "1" | "2" | "3" | "4" | "5" | "6" | "7" | "8" | "9" 
\end{verbatim}
\end{boxedminipage}

The same looking metasymbol is used later as a pure extension point:

\noindent\begin{boxedminipage}{\textwidth}\footnotesize
\begin{verbatim}
<symbol>             ::= <html-unsafe-symbol> | <underscore> | "." | "," | ...
\end{verbatim}
\end{boxedminipage}

The crucial difference in the semantics of these two metasymbols both denoted as ``\texttt{...}'' lies in the fact that in the former one
(i.e., \texttt{"A" | ... | "Z"}) it is basically a macro definition that can be expanded by any human reader, but in the latter one
(i.e., \texttt{"." | ...}) the only thing the reader learns from looking at it is that something can or should be added. Hence, following
the conclusions we drew above, we expand the former omission metasymbol right in the grammar source, but we remove the latter omission metasymbol
with grammar transformation\footnote{Part of \linktoxbgf{remove-extension-points}.}:

\noindent
{\footnotesize
\begin{lstlisting}[language=pp]
vertical( in symbol );
removeV(
 symbol:
        "." "." "."
);
horizontal( in symbol );
\end{lstlisting}}

Finally, we notice some of the productions using explicit right recursion:

\noindent\begin{boxedminipage}{\textwidth}\footnotesize
\begin{verbatim}
<newlines>              ::= <newline> [<newlines>]
<space-tabs>            ::= <space-tab> [<space-tabs>]
<spaces>                ::= <space> [<spaces>]
<decimal-number>        ::= <decimal-digit> [<decimal-number>]
<hex-number>            ::= <hex-digit> [<hex-number>]
\end{verbatim}
\end{boxedminipage}

The deyaccifying transformation steps are straightforward\footnote{Part of \linktoxbgf{deyaccify}.}:

\noindent
{\footnotesize
\begin{lstlisting}[language=pp]
massage(
 newlines?,
 (newlines | EPSILON)
 in newlines);
distribute( in newlines );
vertical( in newlines );
deyaccify(newlines);
massage(
 space-tabs?,
 (space-tabs | EPSILON)
 in space-tabs);
distribute( in space-tabs );
vertical( in space-tabs );
deyaccify(space-tabs);
massage(
 spaces?,
 (spaces | EPSILON)
 in spaces);
distribute( in spaces );
vertical( in spaces );
deyaccify(spaces);
massage(
 decimal-number?,
 (decimal-number | EPSILON)
 in decimal-number);
distribute( in decimal-number );
vertical( in decimal-number );
deyaccify(decimal-number);
massage(
 hex-number?,
 (hex-number | EPSILON)
 in hex-number);
distribute( in hex-number );
vertical( in hex-number );
deyaccify(hex-number);
\end{lstlisting}}

We should specially note here that the form used to define \texttt{spaces} prevented us earlier in \S\ref{magic-links} from using less invasive
grammar transformation operators.  What we ideally want is a transformation that is as semantics preserving as possible\footnote{Part of \linktoxbgf{unify-whitespace}.}:

\noindent
{\footnotesize
\begin{lstlisting}[language=pp]
fold(space);
fold(spaces);
\end{lstlisting}}

It is intentional that these two steps affect the whole grammar. We will return to this issue later in \S\ref{whitespace-beautification}.

There are two views given on \texttt{formatting}: an optimistic one and a realistic one. Since the grammar needs to define the allowed syntax
in a structured way, we scrap the the latter\footnote{Complete listing of \linktoxbgf{remove-duplicates}.}:

\noindent
{\footnotesize
\begin{lstlisting}[language=pp]
removeV(
 formatting:
        apostrophe-jungle
);
eliminate(apostrophe-jungle);
\end{lstlisting}}

The whole section describing Inline HTML was removed from \cite{MW-BNF-Inline-text} prior to extraction because it combines two aspects
that are not intended to be defined with (E)BNF: it defines a different language embedded inside the current one (this can be done in a
clean way by using modules in advanced practical frameworks like Rascal~\cite{Rascal}) and it tries to define rules for automated error fixing
(cf. fault-tolerant parsing, tolerant parsing, etc). It suffices to note here that the metalanguage used in the parts of that section that
were formulated not in plain English, is fascinatingly
different from the parts of the MediaWiki grammar that we have already processed: it uses attributed (parametrised) nonterminals and
postfix modifiers for case (in)sensitivity. The same metasyntax is used in the next section about images, so we do need to find a way to
process chunks like this:

\noindent\begin{boxedminipage}{\textwidth}\footnotesize
\begin{verbatim}
 ImageModeManualThumb  ::= mw("img_manualthumb");
 ImageModeAutoThumb    ::= mw("img_thumbnail");  
 ImageModeFrame        ::= mw("img_frame");      
 ImageModeFrameless    ::= mw("img_frameless");  
                       
/* Default settings: */
 mw("img_manualthumb") ::= "thumbnail=", ImageName | "thumb=", ImageName
 mw("img_thumbnail")   ::= "thumbnail" | "thumb";
 mw("img_frame")       ::= "framed" | "enframed" | "frame";
 mw("img_frameless")   ::= "frameless";
                       
 ImageOtherParameter   ::= ImageParamPage | ImageParamUpright | ImageParamBorder
 ImageParamPage        ::= mw("img_page") 
 ImageParamUpgright    ::= mw("img_upright") 
 ImageParamBorder      ::= mw("img_border")
                       
/* Default settings: */
 mw("img_page")        ::= "page=$1" | "page $1"  ??? (where is this used?)
 mw("img_upright")     ::= "upright" [, ["=",] PositiveInteger]
 mw("img_border")      ::= "border"
\end{verbatim}         
\end{boxedminipage}

We try to list the problems within that grammar fragment:

\begin{itemize}
	\item Parametrised nonterminals are used in a style of function calls.
		This is not completely uncommon to grammarware since the invention of van Wijngaarden grammars~\cite{ALGOL68} and
		attribute grammars~\cite{AG-Genesis}, but unnecessary here.
	\item Some productions end with a terminator symbol ``\texttt{;}'', others don't.
	\item Concatenate metasymbol ``\texttt{,}'' is used rather inconsistently
		(occurs between some metasymbols, doesn't occur between some nonterminal symbols).
	\item Inline comments are given in English without consistent explicit separation from the BNF formulae.
\end{itemize}

The shortest way to overcome these difficulties is to reformat them lexically, unchaining parametrised nonterminals and
appending terminator symbols to productions that did not have them. The result looks like this:

\noindent\begin{boxedminipage}{\textwidth}\footnotesize
\begin{verbatim}
 ImageModeManualThumb ::= "thumbnail=", ImageName | "thumb=", ImageName ;
 ImageModeAutoThumb   ::= "thumbnail" | "thumb";
 ImageModeFrame       ::= "framed" | "enframed" | "frame";
 ImageModeFrameless   ::= "frameless";
                      
 ImageOtherParameter  ::= ImageParamPage | ImageParamUpright | ImageParamBorder
 ImageParamPage       ::= "page=$1" | "page $1"; /* ??? (where is this used?) */
 ImageParamUpgright   ::= "upright" [, ["=",] PositiveInteger]
 ImageParamBorder     ::= "border"
\end{verbatim}
\end{boxedminipage}

One of the fragments fixed in this way contains postfix metasymbols for case insensitivity:

\noindent\begin{boxedminipage}{\textwidth}\footnotesize
\begin{verbatim}
 <behaviourswitch-toc>            ::= "__TOC__"i
 <behaviourswitch-forcetoc>       ::= "__FORCETOC__"i
 <behaviourswitch-notoc>          ::= "__NOTOC__"i
 <behaviourswitch-noeditsection>  ::= "__NOEDITSECTION__"i
 <behaviourswitch-nogallery>      ::= "__NOGALLERY__"i
\end{verbatim}
\end{boxedminipage}

These untypical metasymbols are parsed by Grammar Hunter as separate nonterminals, which we remove by
projection\footnote{Complete listing of \linktoxbgf{remove-postfix-case}.}:

\noindent
{\footnotesize
\begin{lstlisting}[language=pp]
project(
 behaviourswitch-toc:
        "__TOC__" <i>
);
project(
 behaviourswitch-forcetoc:
        "__FORCETOC__" <i>
);
project(
 behaviourswitch-notoc:
        "__NOTOC__" <i>
);
project(
 behaviourswitch-noeditsection:
        "__NOEDITSECTION__" <i>
);
project(
 behaviourswitch-nogallery:
        "__NOGALLERY__" <i>
);
\end{lstlisting}}

There is also a mistake that is easily overlooked unless you analyse top and bottom nonterminals (look at the second option):

\noindent\begin{boxedminipage}{\textwidth}\footnotesize
\begin{verbatim}
 ImageAlignParameter        ::= ImageAlignLeft | ImageAlign|Center |
                                ImageAlignRight | ImageAlignNone
\end{verbatim}
\end{boxedminipage}

This extra unnecessary bar is parsed as a regular choice separator, so we need to fix it this way\footnote{Part of \linktoxbgf{fix-names}.}:

\noindent
{\footnotesize
\begin{lstlisting}[language=pp]
replace(
 (ImageAlign | Center),
 (ImageAlignCenter));
\end{lstlisting}}

The same analysis shows us a fragment in the resulting grammar, which is unconnected because \texttt{ImageOption} does not list it with the others\footnote{Part of \linktoxbgf{connect-grammar}.}:

\noindent
{\footnotesize
\begin{lstlisting}[language=pp]
vertical( in image-option );
addV(
 image-option:
        image-other-parameter
);
horizontal( in image-option );
\end{lstlisting}}

The first and the last productions of the Images subsection contain an explicitly marked nonterminal symbol:

\noindent\begin{boxedminipage}{\textwidth}\footnotesize
\begin{verbatim}
ImageInline                ::= "[[" , "Image:" , PageName, ".",
             ImageExtension, ( { <Pipe>, ImageOption, } ) "]]" ;
Caption                    ::= <inline-text>
\end{verbatim}
\end{boxedminipage}

A production in the middle of the Images subsection and the first production of the Media subsection make inconsistent use of a concatenate symbol:

\noindent\begin{boxedminipage}{\textwidth}\footnotesize
\begin{verbatim}
ImageSizeParameter         ::= PositiveNumber "px" ;
MediaInline            ::=  "[[" , "Media:" , PageName "." MediaExtension "]]" ;
\end{verbatim}
\end{boxedminipage}

And, finally, the last production of the Media subsection contains a wrong defining symbol:

\noindent\begin{boxedminipage}{\textwidth}\footnotesize
\begin{verbatim}
MediaExtension = "ogg" | "wav" ;
\end{verbatim}
\end{boxedminipage}

These three problems were reported and overcome by Grammar Hunter but not solved automatically, because usually there is more that
one way to resolve such issues, and a human intervention is needed to make a choice. After the unified notation is
enforced everywhere, we can extract the grammar and continue recovering it with grammar transformation steps.
It should be noted that Grammar Hunter could not resolve the lack of concatenate symbols, since it starts assuming that the following
symbol is a part of the current one (originally the concatenate symbol was proposed in \cite{ISO-EBNF} in order to allow nonterminal names
contain spaces), but it easily dealt with excessive concatenate symbols because they just virtually insert $\varepsilon$ here and there, which
gets easily normalised.

Back to the rest of the section, we have a fragment with essentially an extension point specified in plain English as the right hand side of
 a production:

\noindent\begin{boxedminipage}{\textwidth}\footnotesize
\begin{verbatim}
GalleryImage             ::=   (to be defined: essentially   foo.jpg[|caption] )
\end{verbatim}
\end{boxedminipage}

We can easily decide to disregard this definition in favour of a really working one\footnote{Part of \linktoxbgf{remove-extension-points}.}:

\noindent
{\footnotesize
\begin{lstlisting}[language=pp]
redefine(
 GalleryImage:
        ImageName ("|" Caption)?
);
\end{lstlisting}}

After analysing top and bottom nonterminals, we easily spot \texttt{unespaced-less-than} being bottom and \texttt{unescaped-less-than} being
top---apparently, they were meant to be one, and the other one is a misspelled variation typically found in big handcrafted grammars. The same issue arises with some other nonterminals, apparently this grammar fragment was typed by someone rather careless at spelling\footnote{Part of \linktoxbgf{fix-names}.}:

\noindent
{\footnotesize
\begin{lstlisting}[language=pp]
unite(unespaced-less-than, unescaped-less-than);
unite(ImageParamUpgright, ImageParamUpright);
unite(ImageValignParameter, ImageVAlignParameter);
\end{lstlisting}}

\subsection{Fundamental elements}\label{funda}

Surprisingly for those who did not look at the text of the Inline Text part, the Fundamental Elements \cite{MW-BNF-Fundamentals} does not contain
any new grammar productions for us, because all of them were encountered within the Inline Text, slightly reordered.


\section{Conclusion}

This section contains the list of imperfections found in the MediaWiki grammar definition.
In the parenthesis we refer to the section in the text that unveils the problem or explains it.

\begin{itemize}
	\item Non-extended Backus-Naur form was claimed to be used (\S\ref{ceci-n'est-pas-une-bnf})
	\item Three different metalanguages used for parts of the grammar (\S\ref{article}, \S\ref{noparse-block}, \S\ref{special-block})
	\item Bulleted-list notation for alternatives is used, both untraditional and inconsistent with other grammar fragments (\S\ref{inline})
	\item Atypical metasymbols used:
		\begin{itemize}
			\item ``\texttt{...?}'' (\S\ref{article-title}) --- not defined, assumed to be an extension point
			\item ``\texttt{(?=EOF)}'' (\S\ref{noparse-block}) --- defined in terms of lookahead symbols
			\item ``\texttt{(}'' and ``\texttt{)}'' (\S\ref{noparse-block}) --- unexpectedly used to denote optionality
			\item ``\texttt{[}'' and ``\texttt{]}'' (\S\ref{noparse-block}) --- unexpectedly used for grouping
			\item ``\texttt{+}'' (\S\ref{magic-links}) --- not defined, assumed to be a plus repetition
			\item ``\texttt{?}'' (\S\ref{magic-links}) --- not defined, assumed to be a postfix optionality
			\item ``\texttt{?()}'' (\S\ref{magic-links}) --- not defined, assumed to be a lookahead assertion
			\item ``\texttt{...}'' (\S\ref{special-block}, \S\ref{inline}) --- omissions due to the lack of knowledge
			\item ``\texttt{...}'' (\S\ref{inline}) --- omissions to denote values from the range of alternatives
			\item ``\texttt{(*}'' and ``\texttt{*)}'' (\S\ref{inline}) --- start and end comment symbols
		\end{itemize}
	\item An undesirable omission/extension point metasymbol was used (\S\ref{article-title}, \S\ref{special-block}, \S\ref{inline})
	\item An undesirable exception metasymbol was used (\S\ref{article})
	\item An attempt to use metasyntax to distinguish between two choice semantics (\S\ref{article})
	\item ``Yaccified'' productions with explicit right recursion (\S\ref{article-title}, \S\ref{inline})
	\item Underused metalanguage functionality: obfuscated ``plus'' repetitions and separator lists (\S\ref{article})
	\item Misspelled nonterminal names w.r.t.\ case: \texttt{WhiteSpaces} vs.\ \texttt{Whitespaces} (\S\ref{article}),
		\texttt{InlineText} vs.\ \texttt{inline-text} (\S\ref{special-block}, \S\ref{inline}), etc
	\item Mistyped nonterminal names: \texttt{unespaced-less-than} vs.\ \texttt{unescaped-less-than} and
	\texttt{ImageParamUpgright} vs.\ \texttt{ImageParamUpright} (\S\ref{inline})
	\item Varying grammar fragment delimiters (\S\ref{noparse-block}, \S\ref{magic-links})
	\item Not marking terminals explicitly with the chosen notation (\S\ref{magic-links}) 
	\item Not marking nonterminals explicitly with the chosen notation (\S\ref{magic-links}) 
	\item Escaping special characters with HTML entities (\S\ref{noparse-block})
	\item Usage of ``regexp format'' to specify title legal characters (\S\ref{links})
	\item Insufficient and excessive escaping within ``regexp format'' (\S\ref{links})
	\item Misleading nonterminal symbol name: plural name for a single character (\S\ref{links}, \S\ref{inline})
	\item Improper omission of the end nonterminal metasymbol (\S\ref{links})
	\item Natural language (examples given in parenthesis) as a part of a BNF production (\S\ref{links})
	\item Inherently ambiguous constructs like $a?^+$ and $a*?$ (\S\ref{noparse-block}, \S\ref{links}, \S\ref{special-block})
	\item Excessive bracketing (\S\ref{article}, \S\ref{special-block})
	\item Unintentionally undefined nonterminals (\S\ref{noparse-block})
	\item Referencing PHP variables like \texttt{Sanitizer::\$wgHtmlEntities}
		and configuration functions like \verb|mw("img_thumbnail")| (\S\ref{inline}, \S\ref{wgHtmlEntities})
\end{itemize}

\section{Finishing touches}\label{beautification}

\input{table}
%

\autoref{F:metrics} shows the progress of several grammar metrics during recovery:
TERM is the number of unique terminal symbols used in the grammar,
VAR is the number of nonterminals defined or referenced there,
PROD is the number of grammar production rules (counting each top alternative in them)~\cite{KraftDuffyMalloy}.
We have already discussed bottom and top nonterminals from \cite{Recovery,DeYacc,PLEX} earlier in \S\ref{article}.
It is known and intuitively understood that high numbers of top and bottom nonterminals indicate unconnected grammar.
In the ideal grammar, only few top nonterminals exist (preferably just one, which is the start symbol) and
only few bottoms (only those that need to be defined elsewhere---lexically or in another language)~\cite{Recovery}.
Thus, our finishing touches mostly involved inspection of the tops and bottoms and their elimination.
The very last step called ``subgrammar'' in \autoref{F:metrics} extracted only the desired start symbol
(\texttt{wiki-page}) and all nonterminals reachable from its definition.

Using the terminology of \cite{Recovery}, in this section we move from a level 1 grammar (i.e., raw extracted one)
to a level 2 grammar (i.e., maximally connected one).

\subsection{Defining special nonterminals}\label{defined-specials}

There is a range of nonterminals used in the MediaWiki grammar that have noticeably specific names (starting and ending with a question sign
or being uppercased): they are not defined by the grammar, but usually the text around their definition is enough for a human reader
to derive the intended semantics and then to specify lacking grammar productions. We also unify the naming convention while doing so
(the final steps of that unification will be present in \S\ref{naming-convention}) and leave some nonterminals undefined (bottom) to serve
connection points to other languages (more of that in \S\ref{wgHtmlEntities})\footnote{Complete listing of \linktoxbgf{define-special-symbols}.}:

\noindent
{\footnotesize
\begin{lstlisting}[language=pp]
vertical( in TableCellParameter );
removeV(
 TableCellParameter:
        ?? HTML cell attributes ??
);
addV(
 TableCellParameter:
        html-cell-attributes
);
horizontal( in TableCellParameter );
vertical( in TableParameters );
removeV(
 TableParameters:
        ?? HTML table attributes ??
);
addV(
 TableParameters:
        html-table-attributes
);
horizontal( in TableParameters );
define(
 FROM_LANGUAGE_FILE:
        "#redirect"
);
inline(FROM_LANGUAGE_FILE);
define(
 STRING_FROM_DB:
        "Wikipedia"
);
inline(STRING_FROM_DB);
define(
 STRING_FROM_CONFIG:
        STR
);
inline(STRING_FROM_CONFIG);
define(
 NS_CATEGORY:
        "Category"
);
inline(NS_CATEGORY);
define(
 ALLOWED_PROTOCOL_FROM_CONFIG:
        "http://"
        "https://"
        "ftp://"
        "ftps://"
        "mailto:"
);
inline(ALLOWED_PROTOCOL_FROM_CONFIG);
unite(LEGAL_ARTICLE_ENTITY, article-title);
\end{lstlisting}}

\subsection{Unification of whitespace and lexicals}\label{whitespace-beautification}

Another big metacategory of nonterminal symbols represent the lexical part, which is not always properly specified by a syntactic
grammar. In the MediaWiki grammar case, there were several attempts to cover all lexical peculiarities including problems arising from
using Unicode (i.e., different types of spaces and newlines), so the least we can do is to unify those attempts. Future work on
deriving a level 3 grammar from the result of this project, will use test-driven correction to complete the lexical part correctly~\cite{Recovery}.
Our current goal is to provide a high quality level 2 grammar without destroying too much information that can be reused later\footnote{Part of \linktoxbgf{unify-whitespace}.}:

\noindent
{\footnotesize
\begin{lstlisting}[language=pp]
unite(??_variants_of_spaces_??, space);
unite(??_carriage_return_and_line_feed_??, newline);
unite(??_carriage_return_??, CR);
unite(??_line_feed_??, LF);
inline(NewLine);
unfold(newline in Whitespaces);
fold(newline in Whitespaces);
unite(??_tab_??, TAB);
\end{lstlisting}}

Another specificity is only referenced but not defined directly by the grammar. According to the text of Inline Text section~\cite{MW-BNF-Inline-text},
this is a patch for dealing with French punctuation. It is highly debatable whether such specificity should be found in the baseline grammar, but since it
is not defined properly anyway, we decide to root it out\footnote{Part of \linktoxbgf{unify-whitespace}.}:

\noindent
{\footnotesize
\begin{lstlisting}[language=pp]
vertical( in text-with-formatting );
removeV(
 text-with-formatting:
        open-guillemet
);
removeV(
 text-with-formatting:
        close-guillemet
);
horizontal( in text-with-formatting );
\end{lstlisting}}

Some bottom lexical nonterminals are trivially defined in BGF\footnote{Part of \linktoxbgf{define-lexicals}.}:

\noindent
{\footnotesize
\begin{lstlisting}[language=pp]
define(
 TAB:
        "\t"
);
define(
 CR:
        "\r"
);
define(
 LF:
        "\n"
);
define(
 any-text:
        unicode-character*
);
define(
 sort-key:
        any-text
);
define(
 any-supported-unicode-character:
        ANY
);
\end{lstlisting}}

\subsection{Connecting the grammar}

The Magic Links part (see \ref{magic-links}) apparently referenced some nonterminals that were never used. We can easily pinpoint them with
a simple grammar analysis showing bottom nonterminals, and after that program the appropriate transformations\footnote{Part of \linktoxbgf{connect-grammar}.}:

\noindent
{\footnotesize
\begin{lstlisting}[language=pp]
define(
 digits:
        digit+
);
unite(digit, decimal-digit);
unite(DIGIT, decimal-digit);
\end{lstlisting}}

Undefined nonterminals \texttt{PositiveInteger} and \texttt{PositiveNumber} both can be merged with this new nonterminal\footnote{Part of \linktoxbgf{connect-grammar}.}:

\noindent
{\footnotesize
\begin{lstlisting}[language=pp]
unite(PositiveInteger, digits);
unite(PositiveNumber, digits);
\end{lstlisting}}

Nonterminal \texttt{newlines} defined at \cite{MW-BNF-Inline-text} and \cite{MW-BNF-Fundamentals}, is also never used and can be eliminated\footnote{Part of \linktoxbgf{connect-grammar}.}:

\noindent
{\footnotesize
\begin{lstlisting}[language=pp]
eliminate(newlines);
\end{lstlisting}}

Last connecting steps are easy since there are not that many top and bottom nonterminals left, and a simple
human inspection can show that some of them are actually misspelled pairs like this one\footnote{Part of \linktoxbgf{connect-grammar}.}:

\noindent
{\footnotesize
\begin{lstlisting}[language=pp]
unite(ImageModeThumb, image-mode-auto-thumb);
unite(category, category-link);
\end{lstlisting}}

In Links section~\cite{MW-BNF-Links} there is a discussion on whether there should be a syntactic category for all links (i.e., internal and
external). The discussion seems to be unfinished, with the nonterminal \texttt{link} specified, but unused (i.e., top). Since the definition
is already available, we decided to use it by folding wherever possible\footnote{Part of \linktoxbgf{connect-grammar}.}:

\noindent
{\footnotesize
\begin{lstlisting}[language=pp]
fold(link);
\end{lstlisting}}

\subsection{Mark exclusion}

BGF does not have a metaconstruct for exclusion (``\texttt{a} should be parseable as \texttt{b} but not as \texttt{c}'', mostly specified as
``\texttt{<a> ::= <b> - <c>}'' within the MediaWiki grammar), but we still want to preserve the information for further refactoring. One of the
ways to do so is to used a marking construct usually found in parameters to transformation operators such as \textbf{project} or
\textbf{addH}\footnote{Part of \linktoxbgf{fake-exclusion}.}:

\noindent
{\footnotesize
\begin{lstlisting}[language=pp]
replace(
 ??_all_supported_Unicode_characters_??_-_Whitespaces,
 <(any-supported-unicode-character Whitespaces)>);
replace(
 UnicodeCharacter_-_WikiMarkupCharacters,
 <(UnicodeCharacter WikiMarkupCharacters)>);
replace(
 SectionLinkCharacter_- "=",
 <(SectionLinkCharacter "=")>);
replace(
 UnicodeCharacter_- "]",
 <(UnicodeCharacter "]")>);
replace(
 UnicodeCharacter_-_BadTitleCharacters,
 <(UnicodeCharacter BadTitleCharacters)>);
replace(
 UnicodeCharacter_-_BadSectionLinkCharacters,
 <(UnicodeCharacter BadSectionLinkCharacters)>);
\end{lstlisting}}

\subsection{Naming convention}\label{naming-convention}

There are three basic problems with the naming convention if we look at the whole extracted grammar, namely:

\begin{description}
	\item[Unintelligible nonterminal names.]
			When looking at a particular grammar production rule situated close to a piece of text explaining all kinds
			of details that did not fit in the BNF, it is easy to overlook non-informative names. In the case of MediaWiki,
			in the final grammar we have bottom nonterminals with the names like
				\verb|FROM_LANGUAGE_FILE|,
				\verb|STRING_FROM_CONFIG|,
				\verb|STRING_FROM_DB|.
			Such names do not belong in the grammar, because they obfuscate it, and the main reason for having a grammar printed
			out in an EBNF-like form in the first place is to make it readable for a human.
	\item[Letters capitalisation.]
			Nonterminal names can be always written in lowercase, or in uppercase, or in any mixture of them.
			The choice of parsing technology can influence that choice: for instance, Rascal~\cite{Rascal} can only
			process capitalised nonterminal names and ANTLR~\cite{ANTLR} treats uppercase nonterminals and non-uppercase ones
			differently. These implicit semantic details need to be acknowledged and accounted for, in a consistent manner,
			which was not the case in the MediaWiki grammar.
	\item[Word separation.]
			Most of the nonterminals have names that consist of several natural words (e.g., ``wiki'' and ``page'').
			There are several ways to separate them:
				by straightforward concatenating (``wikipage''),
				by camelcasing (``WikiPage'' or ``wikiPage''),
				by hyphenating (``wiki-page''),
				by allowing spaces in nonterminal names (``wiki page''),
				etc.
			It does not matter too much which convention is used, as long as it is the same throughout the whole grammar.
			In the case of MediaWiki there is no consistency, which leads to not only decreased readability, but also
			to problems like \texttt{noparse-block} being defined in \cite{MW-BNF-Noparse-block} and
			\texttt{noparseblock} being used in \cite{MW-BNF-Inline-text} (they were obviously meant to be one nonterminal).
\end{description}

The complete transformation script enforcing a consistent naming convention and fixing related problems on the way, looks like this\footnote{Part of \linktoxbgf{fix-names}.}:

\noindent
{\footnotesize
\begin{lstlisting}[language=pp]
unite(noparseblock, noparse-block);
unite(GalleryBlock, gallery-block);
unite(ImageInline, image-FAKE-INLINE);
unite(MediaInline, media-FAKE-INLINE);
unite(Table, table);
unite(Text, text);
unite(InlineText, FAKE-INLINE-text);
unite(Pipe, pipe);
renameN(AnyText, any-text);
renameN(BadSectionLinkCharacters, bad-section-link-characters);
renameN(BadTitleCharacters, bad-title-characters);
renameN(Caption, caption);
renameN(GalleryImage, gallery-image);
renameN(ImageAlignCenter, image-align-center);
renameN(ImageAlignLeft, image-align-left);
renameN(ImageAlignNone, image-align-none);
renameN(ImageAlignParameter, image-align-parameter);
renameN(ImageAlignRight, image-align-right);
renameN(ImageExtension, image-extension);
renameN(ImageModeAutoThumb, image-mode-auto-thumb);
renameN(ImageModeFrame, image-mode-frame);
renameN(ImageModeFrameless, image-mode-frameless);
renameN(ImageModeManualThumb, image-mode-manual-thumb);
renameN(ImageModeParameter, image-mode-parameter);
renameN(ImageName, image-name);
renameN(ImageOption, image-option);
renameN(ImageOtherParameter, image-other-parameter);
renameN(ImageParamBorder, image-param-border);
renameN(ImageParamPage, image-param-page);
renameN(ImageParamUpright, image-param-upright);
renameN(ImageSizeParameter, image-size-parameter);
renameN(ImageValignBaseline, image-valign-baseline);
renameN(ImageValignBottom, image-valign-bottom);
renameN(ImageValignMiddle, image-valign-middle);
renameN(ImageVAlignParameter, image-valign-parameter);
renameN(ImageValignSub, image-valign-sub);
renameN(ImageValignSuper, image-valign-super);
renameN(ImageValignTextBottom, image-valign-text-bottom);
renameN(ImageValignTextTop, image-valign-text-top);
renameN(ImageValignTop, image-valign-top);
renameN(Line, line);
renameN(LinkTitle, link-title);
renameN(MediaExtension, media-extension);
renameN(PageName, page-name);
renameN(PageNameLink, page-name-link);
renameN(PlainText, plain-text);
renameN(SectionLink, section-link);
renameN(SectionLinkCharacter, section-link-character);
renameN(SectionTitle, section-title);
renameN(TableCellParameters, table-cell-parameters);
renameN(TableColumn, table-column);
renameN(TableColumnLine, table-column-line);
renameN(TableColumnMultiLine, table-column-multiline);
renameN(TableFirstRow, table-first-row);
renameN(TableParameters, table-parameters);
renameN(TableRow, table-row);
renameN(TitleCharacter, title-character);
renameN(UnicodeCharacter, unicode-character);
renameN(UnicodeWiki, unicode-wiki);
renameN(WikiMarkupCharacters, wiki-markup-characters);
\end{lstlisting}}

As one can see, we reinforce hyphenation in almost all places, except for nonterminals inherited from other languages (e.g., \texttt{blockquote} from HTML).
The list of plain renamings was derived automatically by a Python one-liner that transformed CamelCase to dash-separated names.
The XBGF engine always checks preconditions for renaming a nonterminal (i.e., the target name must be fresh), so then it was
trivial to turn the non-working \textbf{renameN} calls into \textbf{unite} calls.

\subsection{Embedded languages}\label{wgHtmlEntities}

We may recall seeing \texttt{wgHtmlEntities} undefined nonterminal being referenced in \S\ref{inline}.
There are more like it---in fact, at the end of our recovery project there are 8 bottom nonterminals in the grammar:

\begin{itemize}
	\item \verb|LEGAL_URL_ENTITY|: designates a character that is allowed in a URL; defined by the corresponding RFC~\cite{RFC3986}.
	\item \texttt{inline-html}: was removed deliberately due to incompleteness and questionable representation; defined partially by the accompanying
			English text, partially by the HTML standard~\cite{W3C-HTML}.
	\item \texttt{math-block}: the syntax used by the math extension to MediaWiki~\cite{MW-Math}.
	\item \texttt{CSS}: cascading style sheets used to specify layout of tables and table cells~\cite{W3C-CSS}.
	\item \texttt{html-table-attributes} and \texttt{html-cell-attributes}: also layout of tables and table cells, but in pure HTML.
	\item \texttt{wgHtmlEntities}: one of the HTML entities (``quot'', ``dagger'', ``auml'', etc).
\end{itemize}

They are all essentially different languages that are reused here, but are not exactly a part of wiki syntax.
Some wiki engines may allow for different subsets of HTML and CSS features to be used within their pages,
but conceptually these limitations are import parameters, not complete definitions.
For instance, we could derive a lacking grammar fragment for \texttt{wgHtmlEntities}
by looking at the file \verb|mw_sanitizer.inc| from MediaWiki distribution\footnote{Available as \linktofile{mediawiki.config.wiki}.}:

\noindent\begin{boxedminipage}{\textwidth}\footnotesize
\begin{verbatim}
<wgHtmlEntities> ::= "Aacute" | "aacute" | "Acirc" | "acirc" | "acute" | "AElig"
| "aelig" | "Agrave" | "agrave" | "alefsym" | "Alpha" | "alpha" | "amp" | "and"
| "ang" | "Aring" | "aring" | "asymp" | "Atilde" | "atilde" | "Auml" | "auml"
| "bdquo" | "Beta" | "beta" | "brvbar" | "bull" | "cap" | "Ccedil" | "ccedil"
| "cedil" | "cent" | "Chi" | "chi" | "circ" | "clubs" | "cong" | "copy"
| "crarr" | "cup" | "curren" | "dagger" | "Dagger" | "darr" | "dArr" | "deg"
| "Delta" | "delta" | "diams" | "divide" | "Eacute" | "eacute" | "Ecirc"
| "ecirc" | "Egrave" | "egrave" | "empty" | "emsp" | "ensp" | "Epsilon"
| "epsilon" | "equiv" | "Eta" | "eta" | "ETH" | "eth" | "Euml" | "euml" | "euro"
| "exist" | "fnof" | "forall" | "frac12" | "frac14" | "frac34" | "frasl"
| "Gamma" | "gamma" | "ge" | "gt" | "harr" | "hArr" | "hearts" | "hellip"
| "Iacute" | "iacute" | "Icirc" | "icirc" | "iexcl" | "Igrave" | "igrave"
| "image" | "infin" | "int" | "Iota" | "iota" | "iquest" | "isin" | "Iuml"
| "iuml" | "Kappa" | "kappa" | "Lambda" | "lambda" | "lang" | "laquo" | "larr"
| "lArr" | "lceil" | "ldquo" | "le" | "lfloor" | "lowast" | "loz" | "lrm"
| "lsaquo" | "lsquo" | "lt" | "macr" | "mdash" | "micro" | "middot" | "minus"
| "Mu" | "mu" | "nabla" | "nbsp" | "ndash" | "ne" | "ni" | "not" | "notin"
| "nsub" | "Ntilde" | "ntilde" | "Nu" | "nu" | "Oacute" | "oacute" | "Ocirc"
| "ocirc" | "OElig" | "oelig" | "Ograve" | "ograve" | "oline" | "Omega"
| "omega" | "Omicron" | "omicron" | "oplus" | "or" | "ordf" | "ordm" | "Oslash"
| "oslash" | "Otilde" | "otilde" | "otimes" | "Ouml" | "ouml" | "para" | "part"
| "permil" | "perp" | "Phi" | "phi" | "Pi" | "pi" | "piv" | "plusmn" | "pound"
| "prime" | "Prime" | "prod" | "prop" | "Psi" | "psi" | "quot" | "radic"
| "rang" | "raquo" | "rarr" | "rArr" | "rceil" | "rdquo" | "real" | "reg"
| "rfloor" | "Rho" | "rho" | "rlm" | "rsaquo" | "rsquo" | "sbquo" | "Scaron"
| "scaron" | "sdot" | "sect" | "shy" | "Sigma" | "sigma" | "sigmaf" | "sim"
| "spades" | "sub" | "sube" | "sum" | "sup" | "sup1" | "sup2" | "sup3" | "supe"
| "szlig" | "Tau" | "tau" | "there4" | "Theta" | "theta" | "thetasym" | "thinsp"
| "THORN" | "thorn" | "tilde" | "times" | "trade" | "Uacute" | "uacute" | "uarr"
| "uArr" | "Ucirc" | "ucirc" | "Ugrave" | "ugrave" | "uml" | "upsih" | "Upsilon"
| "upsilon" | "Uuml" | "uuml" | "weierp" | "Xi" | "xi" | "Yacute" | "yacute"
| "yen" | "Yuml" | "yuml" | "Zeta" | "zeta" | "zwj" | "zwnj"
\end{verbatim}
\end{boxedminipage}

These are 252 entities taken from the DTD of HTML 4.0~\cite{W3C-HTML}. XHTML 1.0 defines an additional entity called
``apos''~\cite{W3C-XHTML}, which, technically speaking, can be handled by MediaWiki since in its current state it
rewrites wikitext to XHTML 1.0 Transitional. Whether it is the grammar's role to report an error when it is used,
remains an open question. Furthermore, suppose we are developing wikiware which is not a WYSIWYG editor, but a
migration tool or an analysis tool: this would mean that the details about all particular entities are of little importance,
and one could define an entity name to be just any alphanumeric word. Questions like these arise when languages are combined,
and for this particular project we leave the bottom nonterminals that represent import points, undefined.

\newpage\section{Results and future work}

This document has reported on a successful grammar recovery effort. The input for this project was a community-created MediaWiki grammar
manually extracted from the PHP tool that is used to transform wiki text to HTML. This grammar contained unconnected fragments in
at least five different notations, bearing various kinds of errors from conceptual underuse of
base notation to simple misspellings, rendering the grammar fairly useless. As an output we provide a level 2 grammar, ready to be
connected to adjacent modules (grammars of HTML, CSS, etc) and made into a higher level grammar (e.g., test it on a real wiki code).
Naturally, this effort is one step in a long way, and we take the rest of the report to sketch the next milestones and planned deliverables:

\begin{description}
	\item[Fix grammar fragments.]
		The first thing we can do is regenerate the original grammar fragments in the same notation.
		One one hand, this would help to not alienate the grammar from its creators; on the other hand,
		the fragments will use a consistent notation throughout the grammar and be validated as
		not having any misspellings, metasymbol omissions, etc.
	\item[Derive several versions.]
		Just in case the same MediaWiki grammar is needed in several different notations (e.g., BNF and EBNF),
		we can derive them from the baseline grammar with either inferred or programmable grammar transformation.
	\item[Propose a better notation.]
		Whether or not the pure BNF grammar is delivered to Wikimedia Foundation, it will be of limited use to most people.
		ANTLR notation that Wiki Creole used, is more useful, but even less easy to comprehend.
		Both more readable and more expressive variants of grammar definition formalisms exist and can be advised for use
		based on the required functionality.
	\item[Find ambiguities and other problems.]
		Various grammar analysis techniques referenced in the text above can be used to perform deeper analyses
		on the grammar in order to make it fully operational in Rascal, resolve existing ambiguities, and perhaps
		even spot problems that are unavoidable with the current notation.
	\item[Complete the lexical part.]
		Some lexical definitions were already found in the source grammar, and were mostly preserved through the recovery
		process. A level 3 grammar can be derived from our current result by reinspecting these definitions together with
		textual annotations found on MediaWiki.org.
\end{description}
%

\newpage
\bibliographystyle{abbrv}
\bibliography{paper}

\end{document}

%% file: table.tex
\begin{table}[t]\footnotesize
\begin{center}
\begin{tabular}{|l|c|c|c|c|c|}\hline
												& TERM & VAR & PROD & Bottom & Top \\\hline
	After extraction							&  304 & 188 &  691 &     78 &  29 \\
	After \linktoxbgf{utilise-repetition}		&  304 & 188 &  691 &     78 &  29 \\
	After \linktoxbgf{remove-concatenation}		&  304 & 188 &  691 &     78 &  29 \\
	After \linktoxbgf{remove-extension-points}	&  304 & 188 &  684 &     73 &  29 \\
	After \linktoxbgf{remove-php-legacy}		&  302 & 188 &  684 &     70 &  29 \\
	After \linktoxbgf{deyaccify}				&  302 & 187 &  680 &     70 &  29 \\
	After \linktoxbgf{remove-comments}			&  300 & 187 &  680 &     68 &  29 \\
	After \linktoxbgf{remove-lookahead}			&  300 & 184 &  680 &     66 &  29 \\
	After \linktoxbgf{remove-duplicates}		&  300 & 183 &  678 &     66 &  29 \\
	After \linktoxbgf{dehtmlify}				&  299 & 183 &  678 &     66 &  29 \\
	
	After \linktoxbgf{utilise-question}			&  299 & 183 &  678 &     66 &  29 \\
	After \linktoxbgf{fix-markup}				&  299 & 183 &  678 &     64 &  29 \\
	After \linktoxbgf{define-special-symbols}	&  299 & 183 &  678 &     62 &  29 \\
	After \linktoxbgf{fake-exclusion}			&  299 & 183 &  678 &     58 &  26 \\
	After \linktoxbgf{remove-postfix-case}		&  299 & 183 &  678 &     57 &  26 \\
	After \linktoxbgf{fix-names}				&  307 & 182 &  681 &     37 &  14 \\
	After \linktoxbgf{unify-whitespace}			&  307 & 181 &  681 &     31 &  13 \\
	After \linktoxbgf{connect-grammar}			&  307 & 181 &  671 &     16 &   7 \\
	After \linktoxbgf{refactor-repetition}		&  307 & 181 &  671 &     16 &   7 \\
	After \linktoxbgf{define-lexicals}			&  310 & 187 &  671 &      9 &   7 \\
	After subgrammar							&  310 & 177 &  664 &      8 &   1 \\\hline
\end{tabular}
\end{center}
\caption{Simple metrics computed on grammars during transformation.}
\label{F:metrics}
\end{table}